\documentclass{aa}
\usepackage{graphicx, natbib, longtable, lscape, amssymb}
\usepackage[english]{babel}
\usepackage{lscape}

\newcommand{\gppr}{\stackrel{>}{\scriptstyle \sim}}
\newcommand{\gappr}{\raisebox{-0.4ex}{$\gppr$}}
\newcommand{\lppr}{\stackrel{<}{\scriptstyle \sim}}
\newcommand{\lappr}{\raisebox{-0.4ex}{$\lppr$}}
\newcommand{\Porb}{\mbox{$P_\mathrm{orb}$}}

\newcommand{\ali}{\mbox{$\alpha_\mathrm{int}$}}

\newcommand{\Mwd}{\mbox{$M_\mathrm{wd}$}}
\newcommand{\Rwd}{\mbox{$R_\mathrm{wd}$}}

\newcommand{\Msun}{\mbox{$\mathrm{M}_{\odot}$}}
\newcommand{\Rsun}{\mbox{$R_{\odot}$}}
\newcommand{\Teff}{\mbox{$T_\mathrm{eff}$}}

\newcommand{\Lacc}{\mbox{$L_\mathrm{acc}$}}

\begin{document}

\title{Post-common-envelope binaries from SDSS. XI: \\
The white dwarf mass distributions of CVs and pre-CVs}
\titlerunning{White dwarf mass distributions of CVs and pre-CVs}
\author{M. Zorotovic\inst{1,2,3}, M.R. Schreiber\inst{1}, B.T. G\"ansicke\inst{4}}
\authorrunning{Zorotovic et al.}
\institute{Departamento de F\'isica y Astronom\'ia, Facultad de Ciencias, Universidad de
Valpara\'iso, Valpara\'iso, Chile 
\email{mzorotov@astro.puc.cl}
\and
Departamento de Astronom\'ia y Astrof\'isica, Pontificia Universidad Cat\'olica de Chile, Santiago, Chile 
\and
European Southern Observatory, Alonso de Cordova 3107, Santiago, Chile
\and
Department of Physics, University of Warwick, Coventry CV4 7AL, UK
}
\offprints{M. Zorotovic}

\date{Received 2 February 2011 / Accepted 2 August 2011}

\abstract {We have known for a long time that many of the measured WD masses in cataclysmic variables (CVs) significantly exceed the mean mass of single white dwarfs (WDs). This was thought to be related to observational biases, but recent high-precision measurements of WD masses in a great number of CVs are challenging this interpretation. A crucial question in this context is whether the high WD masses seen among CVs are already imprinted in the mass distribution of their progenitors, i.e. among detached post-common-envelope binaries (PCEBs) that consist of a WD and a main-sequence star.}
{We review the measured WD masses of CVs, determine the WD-mass distribution of an extensive sample of PCEBs that are representative for the progenitors of the current CV population (\textit{pre-CVs}) and compare both distributions.}
{We calculate the CV formation time of the PCEBs in our sample by determining the post common-envelope (CE) and the main-sequence evolution of the binary systems and define a pre-CV to be a PCEB that evolves into a semi-detached configuration with stable mass transfer within less than the age of the Galaxy. Possible observational biases affecting the WD-mass distribution for the pre-CV and the CV samples are discussed.}
{The mean WD mass among CVs is $\langle\Mwd\rangle=0.83\pm0.23$\thanks{The $\pm$ uncertainty refers to the $1 \sigma$ standard deviation of the mass distribution. All uncertainties attached to mean masses quoted throughout this paper are computed this way and do not refer to the standard
error of the mean (which is smaller by a factor of $1/\sqrt{N}$).}\,\Msun, much larger than that found for pre-CVs, $\langle\Mwd\rangle=0.67\pm0.21$\,\Msun. Selection effects cannot explain the high WD masses observed in CVs. We also note that compared to the predictions of binary-population models, the observed fraction of He-core WDs is small both among CVs ($\leq10\%$) and pre-CVs ($\leq 17 \pm 8\%$). }
{We suggest two possible explanations for the high WD masses among CVs, both of which imply substantial revisions to the standard model of CV evolution: either most CVs have formed above the orbital-period gap (which requires a high WD mass to initiate stable mass transfer or a previous phase of thermal-timescale mass transfer), or the mass of the WDs in CVs grows through accretion (which strongly disagrees with the predictions of classical nova models). Both options may imply that CVs contribute to the single-degenerate progenitors of Type Ia supernovae. The number of He-core WDs in CVs ($\leq10\%$) is roughly consistent with the number of He-core WDs in pre-CVs ($\leq17\pm8\%$). This indicates a low value of the CE efficiency. }


\maketitle

\section{Introduction} \label{sec:intro}

Cataclysmic variables (CVs) are compact binaries containing a white dwarf (WD) that accretes mass from a Roche-lobe filling donor via stable mass transfer \citep[see][ for an encyclopedic review]{warner95-1}. The standard scenario for the formation of CVs proposed by \citet{paczynski76-1} assumes that the progenitor systems were relatively wide binaries in which the more massive star (the primary) fills its Roche lobe during the first giant branch (FGB) or asymptotic giant branch (AGB) phase.  Then, dynamically-unstable mass transfer from the giant to the less massive companion leads to a common-envelope (CE) configuration with the giant's envelope engulfing the future WD and its companion. Owing to friction within the envelope, angular momentum and orbital energy are transferred from the binary orbit to the CE, and the separation between the primary core and the secondary star decreases in a spiralling-in process until the envelope is expelled, which is thought to happen in $\lappr\,10^3$\,yr \citep{taam+sandquist00-1}. The CE evolution terminates the mass growth of the core of the primary and produces a short-orbital-period detached post-common-envelope binary (PCEB) consisting of the core of the primary (typically a WD) and a low-mass secondary main-sequence (MS) star. The PCEBs eventually evolve into CVs through orbital-angular-momentum loss by gravitational radiation and magnetic-wind braking.

Since the first systematic studies of CVs were carried out, the mean masses of WDs in CVs have been found to lie in the range of $\langle\Mwd\rangle=0.8-1.2\Msun$, \citep{warner73-1, ritter76-1, warner76-1, robinson76-1, ritter87-1}, i.e. significantly exceeding the average mass of single WDs of $\langle\Mwd\rangle\simeq0.6\Msun$ \citep[e.g.][]{koesteretal79-1, bergeronetal92-1, kepleretal07-1}. This finding strongly contradicts the expectations for the intrinsic distribution of WD masses in CVs. Three evolutionary processes affect the WD masses in CVs: (1) CE evolution terminates the core growth of the primary stars, which reduces the expected CV WD masses compared to single WDs. (2) The second phase of mass transfer must be stable, which allows for a broader range of secondary masses for systems containing high-mass WDs. (3) Classical nova eruptions are supposed to slowly erode the WD masses of CVs \citep{prialnik86-1,prialnik+kovetz95-1}, which should lower the WD masses of CVs.  Processes (1) and (3) will lower the average WD mass, (2) will increase it. According to the most frequently assumed initial-mass-ratio distributions, initial-primary-mass function, and CE efficiency, CE truncation and WD erosion caused by nova eruptions should largely compensate the stability argument and the intrinsic CV WD mass distribution is expected to be dominated by WDs with $\Mwd\,\lappr\,0.6$\,\Msun.  In particular, depending on the assumed CE efficiency, up to $50\%$ of CV primaries should be He-core WDs with $\Mwd\,\lappr\,0.47$\,\Msun \citep[e.g.][]{dekool92-1,politano96-1} but very few, if any, candidates have so far been identified \citep[see e.g.][]{shenetal09-1}.

The measured high mean CV WD masses and the nearly complete absence of He-core WDs in observed samples have  been explained as a selection effect by \citet{ritter+burkert86-1}. Assuming that CVs are in general discovered because of the accretion generated luminosity, which is a strong function of the WD mass ($\Lacc \propto \Mwd/\Rwd$, with  $\Rwd \propto \Mwd^{-\alpha}$), these authors showed that this would lead to an observational bias in any magnitude-limited CV sample. For $V<12.5$, \citet{ritter+burkert86-1} found that standard CV-population models would result in an \textit{observed} mean WD mass of $\simeq0.9$\,\Msun, consistent with the observations available at that time. For fainter CV samples, the authors predicted a decreasing mean WD mass, down to $\simeq0.66$\,\Msun\ for $V<20$. In addition, the bias considered by \citet{ritter+burkert86-1} should be much stronger above the orbital-period gap than below the gap, which should cause the mean WD mass of observed samples to be significantly higher above than below the gap.

Thanks to the Sloan Digital Sky Survey \citep[SDSS][]{adelman-mccarthyetal08-1, abazajianetal09-1}, we now have a large, homogeneous sample of faint ($i\leq19.1$) CVs (\citealt{szkodyetal09-1}, and references therein) that contains a substantial number of short-period CVs accreting at very low rates \citep{gaensickeetal09-2}. If the high masses of CVs in the early samples were indeed related to a selection effect related to their higher luminosity \citep{ritter+burkert86-1}, this effect should be much less pronounced among the SDSS CVs.  The measured WD masses of SDSS CVs, however, defy this expectation, because they are just as high as those among bright CVs \citep[e.g.][]{littlefairetal06-2,littlefairetal08-1,savouryetal11-1}. It is thus time to re-assess the WD-mass distribution of CVs, and discuss the findings in the context of CV evolution. 

Important in this respect, SDSS has not only provided a deep sample of CVs, but also the first-ever large homogeneous population of detached white-dwarf/main-sequence (WDMS) binaries \citep[e.g.][]{silvestrietal07-1, rebassa-mansergasetal10-1} of which a considerable fraction, $\sim30$\%, are PCEBs (\citealt{rebassa-mansergasetal07-1, rebassa-mansergasetal08-1, rebassa-mansergasetal10-1, schreiberetal08-1, nebot-gomez-moranetal09-1, pyrzasetal09-1, schwopeetal09-1} and Nebot G\'omez-Mor\'an et al. 2011, submitted). Those PCEBs that will evolve into semi-detached binaries undergoing stable mass transfer on timescales shorter than the age of the Galaxy are representative of the progenitors of the current-day CV population.

We determine and compare the mass distributions of WDs in CVs and the WDs in their progenitors, investigate potential observational biases, and discuss potential causes for the the high masses of CV WDs. We start with a brief review of mass determinations of CV WDs.

\section{The WD masses of CVs}\label{s:cvwd-masses}

A variety of methods can be used to determine the masses of WDs in CVs. In principle, knowledge of the radial velocity amplitudes of both stars, $K_1$ and $K_2$, along with the orbital period and binary inclination, allows one to calculate both stellar masses straight away using Kepler's third law. Knowledge regarding the inclination of the system usually comes from modelling its light curve, with the most accurate values being derived for eclipsing CVs.  However, in most CVs, this method is only of limited use, because the WDs (and/or the companion stars) are outshone at optical wavelengths by the accretion disc/stream. Even if the WD (companion star) is sufficiently hot to dominate at ultraviolet (infrared) wavelengths, technological limitations have so far prevented dynamical mass measurements at wavelengths other than the optical.

Hence, much of the published work has been based on radial velocity proxies for either the WD or the companion star. For the WD, one commonly made assumption is that the emission lines of the accretion disc follow the motion of the WD around the centre of mass of the binary, and can hence be used to infer the radial velocity amplitude of the WD. However, in many cases the radial velocity curves of the emission lines are asymmetric, and/or offset in phase with respect to the expected motion of the WD, casting some doubt on the underlying assumption of this method. The pitfalls and limitations of dynamical mass determinations in CVs are discussed in detail by \citet{shafter83-1} and \citet{thorstensen00-1}.

Modelling the optical light curves of eclipsing CVs where both WD and bright spot are detected allows one to constrain the stellar masses and radii, based on the assumptions that the bright spot lies along the ballistic trajectory of the accretion stream leaving the secondary star, and a WD mass-radius relation. The concept was pioneered for Z\,Cha by \citet{smak79-1} and \citet{cook+warner84-1}, and subsequently adopted for a handful of bright eclipsing CVs. The advent of ULTRACAM on large aperture telescopes \citep{dhillonetal07-1}, and an increasing sample of eclipsing CVs, predominantly from the SDSS \citep{szkodyetal09-1}, led to a significant increase in the number of CVs with published high-precision WD masses \citep[e.g.][]{felineetal04-2, littlefairetal06-2, savouryetal11-1}.

Another fairly robust method becomes available if the radial velocity amplitudes as well as the systemic ($\gamma$-) velocities of both stars can be measured. The difference between the $\gamma$-velocities of both stars corresponds to the gravitation redshift of the WD, which is a measure of its surface gravity, $g=G\Mwd/\Rwd^2$. An additional input for this method is a mass-radius relation for the WD. An application to U\,Gem \citep{sionetal98-1, long+gilliland99-1} finds a high mass for the WD, consistent with earlier estimates based on eclipse/dynamical studies \citep{zhangetal87-1}.

Finally, the WD mass can also be inferred from ultraviolet spectroscopy if it dominates the emission from the CV at those wavelengths and has an accurate distance measurement. The ultraviolet spectrum is fitted for the effective temperature over a fixed range of WD masses. The scaling factor between the model flux and observed flux then yields the WD radius, which, combined with a WD mass-radius relation, yields a unique solution for the WD mass. See \cite{longetal04-1} and \citet{gaensickeetal06-2} for applications of this method to WZ\,Sge and AM\,Her.

\subsection{Observational census}\label{sec:cvmass}

\begin{figure}
\centering
\includegraphics[angle=270,width=0.49\textwidth]{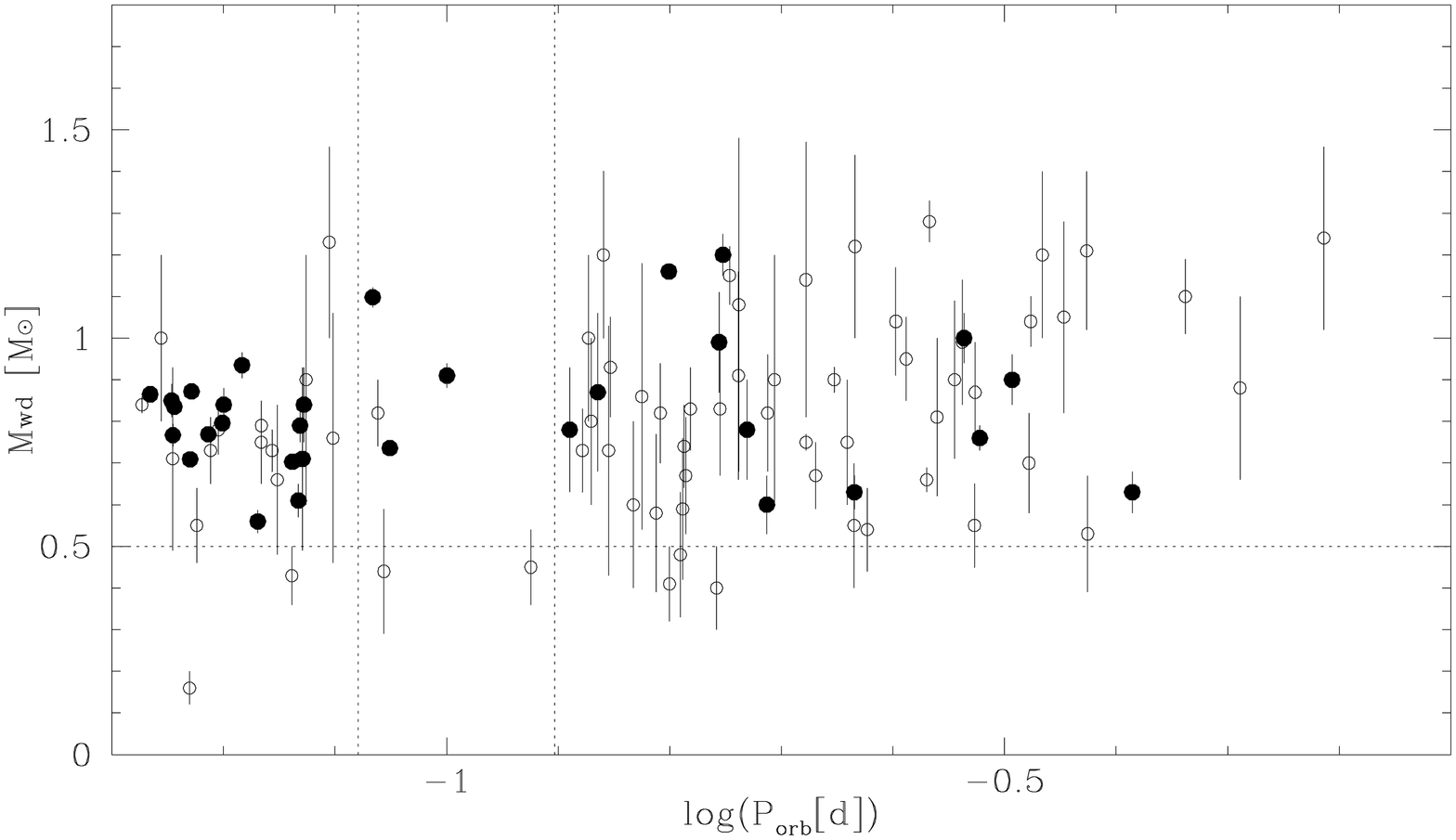}
\caption{CV WD masses from \citet[][V7.14]{ritter+kolb03-1} and 
  \citet{savouryetal11-1}, excluding
  systems with no published uncertainty in $\Mwd$, and GK\,Per, for
  which only a lower limit $\Mwd\ge0.87\pm0.24$\,\Msun\ has been
  determined. 
  Systems shown as solid symbols are our ``fiducial''
  sub-sample (Table\,\ref{tab:cvmass}). The dotted horizontal line
  indicates the division between C/O-core WDs (above) and He-core WDs
  (below), while the dotted vertical lines show the position of the
  orbital-period gap.
  }
\label{fig:cvmass}
\end{figure}

The observational challenge in measuring CV WD masses is underlined by
the fact that only 116 out of 849 CVs listed in the catalogue of
\citet[][V7.14]{ritter+kolb03-1} have an entry for their WD
mass. Figure\,\ref{fig:cvmass} shows the orbital-period distribution
of these WD masses, where we exclude systems with no published
uncertainty of the WD mass, as well as a lower limit for the mass of
the WD in GK\,Per, and we include the recent results of
  \citet{savouryetal11-1}.  The resulting final sample contains
104 objects. From this list of systems we defined a
``fiducial'' sub-sample of 32 robust high-quality WD-mass
measurements (Table\,\ref{tab:cvmass}). This
selection is necessarily subject to some personal bias, but we felt
that comparing a well-defined sub-sample was helpful in assessing the
overall quality of CV WD mass measurements. The mean mass of
the 32 fiducial CV WDs is
$\langle\Mwd\rangle=0.82\pm0.15$\,\Msun, compared to
$0.83\pm0.23$\,\Msun\ for the whole sample, i.e. both samples
are fully consistent with the major difference being the smaller
standard deviation among the fiducial CVs.  Inspecting
Fig.\,\ref{fig:cvmass} strongly suggests that this standard deviation
is related to intrinsic scatter in the CV WD masses, rather than to
the uncertainties in the measurements. The mean of the
fiducial CV WD masses below and above the period gap are
$\langle\Mwd(\Porb\le3\mathrm{\,h})\rangle = 0.80\pm0.12$\,\Msun\ and
$\langle\Mwd(\Porb>3\mathrm{\,h})\rangle = 0.86\pm0.20$\,\Msun,
i.e. there is no significant difference in the mean WD mass between
these two subsets of systems, but the scatter around the mean value is
significantly larger above the gap. That there is no
  significant difference in the mean WD mass above and below the gap
  clearly disagrees with \citeauthor{ritter+burkert86-1}'s (1986)
  prediction of a strong dependence of \Mwd\ on the orbital period.

\addtocounter{table}{1}

The only CV with a WD mass that is significantly below 0.5\,\Msun\ is
T\,Leo (aka QZ\,Vir): \citet{ritter+kolb03-1} quote an extremely low
value of $\Mwd=0.14\pm0.04$\,\Msun\ from the PhD thesis of Allen
Shafter; however, \citet{shafter+szkody84-1} state that \textit{``The mass
  of the WD most likely lies in the range 0.35--0.4\,\Msun.''}  Only
$7 \pm 3\%$ of the 104 CVs with available WD-mass
estimates and errors have $M_1 \leq 0.5$\,\Msun, and none of the
systems in the sub-sample of 32 with presumably more reliable
mass determinations. We therefore conclude that the fraction
  of He-core WDs in the observed sample of CVs is $\leq10\%$.

\subsection{Possible observational biases}\label{sec:cvbias}

Given that the presently observed accretion luminosity of a CV scales
as $\Lacc\propto \Mwd/\Rwd$, and $\Rwd\propto\Mwd^{-\alpha}$ for WDs,
a higher WD mass will result in a higher accretion luminosity, and
hence, at face value, a higher probability of detecting the system in
a magnitude-limited sample. Possible selection effects related to the
WD mass were discussed in some detail by \citet{ritter+burkert86-1}.
Using a simple model for the evolution of CVs and considering a
magnitude-limited sample with a limiting magnitude of $V=10-12.5$, they
find observational selection effects to be sufficient to
explain the observed mean WD masses of $\sim\,0.9-1.1$\,\Msun\,found
for CVs dominated by accretion disc luminosity, i.e. dwarf novae in
outburst and nova-like variables. \citet{ritter+burkert86-1} further
predict that the selection should be weaker for short-orbital-period
systems and that selection effects should nearly disappear for
limiting magnitudes of $V\sim19-20$.

Over the past decade, a great number of short-period CVs in which the
WD dominates the optical emission have been discovered in the SDSS
\citep{szkodyetal09-1,gaensickeetal09-2} and the eclipsing systems are
prime candidates for detailed stellar parameter studies
\citep[e.g.][]{littlefairetal06-2,littlefairetal08-1}. In
  contrast to previous samples, the SDSS CV population was not
selected by outburst properties or X-ray emission, but purely on the
  basis of non-stellar colours. The $ugriz$ colour space spanned by
  the CV population overlaps with that of quasars, and hence the
  completeness of the SDSS CV sample is similar to that of the
  quasars, i.e. $\sim95$\% \citep{richardsetal02-1, richardsetal04-1}.
The brightness of these systems is predominantly set by the luminosity
of the accretion (compressionally) heated WD, and hence depends on the
secular mean accretion rate \citep{townsley+bildsten02-2,
  townsley+bildsten03-1, townsley+bildsten04-1,
  townsley+gaensicke09-1}. This possible mass-dependent observational
bias is not covered by the discussion of \citet{ritter+burkert86-1},
and hence needs to be investigated.

\begin{figure}
\centering
\includegraphics[width=0.49\textwidth]{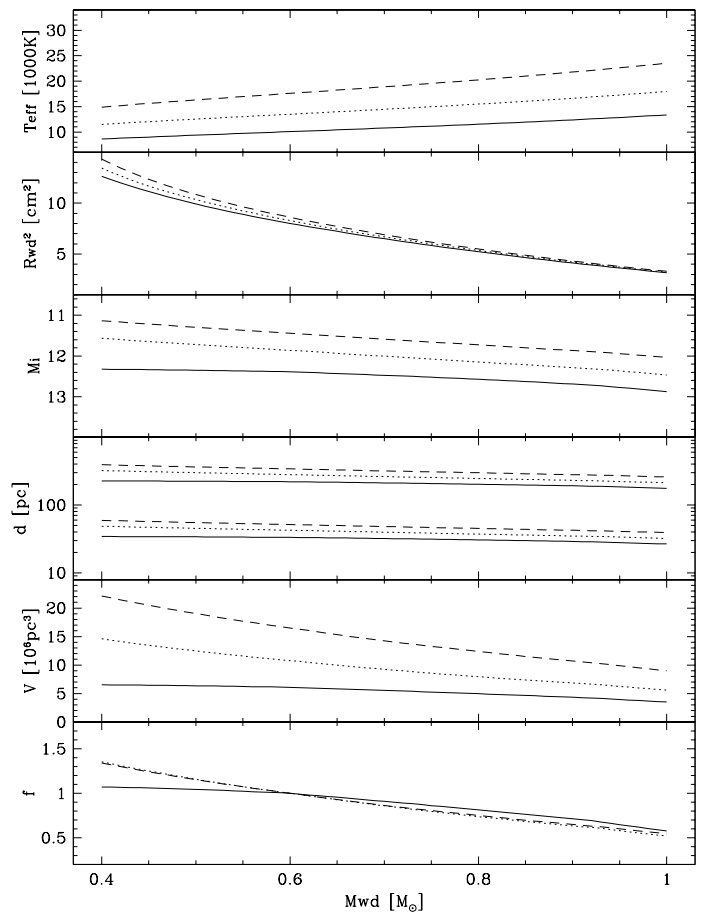}
\caption{The probability of identifying CVs within
  SDSS if their optical emission is dominated by the luminosity of the
  accretion-heated WD as a function of \Mwd. From top to bottom: the
  effective temperature calculated using Eq.\,(1) in
  \citet{townsley+gaensicke09-1}; $\Rwd^2$; the absolute $i$-band
  magnitude; the minimum and maximum distance at which SDSS will have
  obtained follow-up spectroscopy; the effective survey volume for a
  spherical cap with $b>30^{\circ}$, $H_\mathrm{z}=200$\,pc, and
  $d_\mathrm{min}<d<d_\mathrm{max}$, and the relative probability
  normalized to $\Mwd=0.6$\,\Msun. Three secular mean accretion rates
  are shown, $\langle\dot M\rangle=3\times10^{-11}\mathrm{\Msun\,yr^{-1}}$ (solid
  lines), $10^{-10}\mathrm{\Msun\,yr^{-1}}$ (dotted lines), and
  $3\times10^{-10}\mathrm{\Msun\,yr^{-1}}$ (dashed lines).}
\label{fig:cvbias}
\end{figure}

The quiescent WD luminosity $L_\mathrm{q}$ as a function of the
secular mean accretion rate $\langle\dot M\rangle$ and WD mass is given by
Eq.\,(1) in \citet{townsley+gaensicke09-1}. From this,
$\Teff=(L_\mathrm{q}/4\pi\sigma\Rwd^2)^{1/4}$ (Fig.\,\ref{fig:cvbias},
top panel). Given that $L_\mathrm{q}$ only mildly depends on \Mwd, the
increase of \Teff\ as a function of \Mwd\ almost entirely arises
because of the decrease in $\Rwd^2$ (Fig.\,\ref{fig:cvbias}, second
panel form the top). A slight complication is that \Rwd\ depends on
both \Mwd\ and \Teff, and hence Eq.\,(1) from
\citet{townsley+gaensicke09-1} cannot be solved analytically. We
account for this by computing \Teff\ with an initial \Rwd\ using
\citeauthor{nauenberg72-1}'s (1972) zero-temperature mass-radius
relation, and then iteratively re-compute \Teff\ with \Rwd\ from
\citet{bergeronetal95-2}\footnote{Updated cooling models are available
  at
  http://www.astro.umontreal.ca/$\sim$bergeron/CoolingModels/}. \Mwd\ and
\Teff\ determine the absolute $i$-band magnitude, $M_i$, which is
again obtained from the cooling models of \citet{bergeronetal95-2}
(Fig.\,\ref{fig:cvbias}, third panels from the top).

The WD-dominated SDSS CVs were predominantly found as a by-product of
the main (ultraviolet-excess) quasar survey \citep{richardsetal02-1},
which spans a magnitude range $15.5\la i\le19.1$\footnote{Technically,
  the limiting magnitude of the main quasar survey is a
  \textit{de-reddened} $i=19.1$, however, given that extinction is low
  at high galactic latitudes and weak at long wavelengths, this
  introduces only minute changes as a function of the specific
  line-of-sight. The bright end of the spectroscopic follow-up is
  limited by the SDSS imaging saturating.}, and is $\sim95$\% complete
\citep{richardsetal04-1}. $M_i$ determined above therefore translates into
a minimum and maximum distance at which a system will be
spectroscopically followed-up by SDSS (Fig.\,\ref{fig:cvbias}, fourth
panels from the top). The \textit{effective} survey volume of SDSS as
a function of \Mwd\ and $\langle\dot M\rangle$ is then given by integration 
of a spherical cap, adopting galactic latitudes $b>30^{\circ}$ and a scale
height of $H_z=200$\,pc, for distances $d_\mathrm{min}\le d\le
d_\mathrm{max}$ (Fig.\,\ref{fig:cvbias}, fifth panel from the
top). The probability of identifying a CV with a given $\Mwd$ and
$\langle\dot M\rangle$ is proportional to the effective survey volume, and
given, relative to the probability of identifying a CV containing a WD of
$\Mwd=0.6$\,\Msun, in the bottom panel of Fig.\,\ref{fig:cvbias}.

While the probability of identifying WD-dominated CVs in SDSS does not
depend very strongly on \Mwd, a clear trend disfavouring more massive
WDs is observed. This can be understood qualitatively as follows. For
the considered range of \Mwd\ and $\langle\dot M\rangle$, the WD spectrum
in the $i$-band is close to a Rayleigh-Jeans distribution, and hence
the flux $f_i\propto T$. As outlined above, $L_\mathrm{q}$ is only a
mild function of \Mwd, and as $\Teff\propto(L/\Rwd^2)^{1/4}$, the
WD mass-radius relation dominates, resulting in an increase of $M_i$
as a function of \Mwd. Two caveats to this investigation are that
$L_q$ is the average luminosity over a classical nova cycle, the
luminosity of individual systems will scatter somewhat around that
value \citep{townsley+bildsten04-1}, and the bright spot will
contribute some fraction of the $i$-band light, which will lead to a
slight compensation of the bias against high-mass WDs. 

We can summarize our findings for the WD masses in CVs as follows:
\begin{itemize}
\item The mean mass of CV WDs is $\simeq0.8\Msun$, independent
  of their apparent magnitude.
\item The mean WD masses of CVs below and above the
  orbital-period gap do not differ significantly.
\item For short-period WD-dominated CVs, similar to those
  identified by the SDSS, the observed high average mass
  (Sect.\,\ref{sec:cvmass}) has to be taken as a good representation
  of the intrinsic mass distribution of these systems; in particular,
  there is no significant bias against finding low-mass/He-core WDs.
\end{itemize}
 Unless the recent high-quality measurements of CV WD
  masses are systematically wrong, these three findings are in clear
  contrast to the predictions of \citet{ritter+burkert86-1}, i.e. that
  the average WD mass within an observed sample of CVs should decrease
  with increasing limiting magnitude, and that the average WD mass should
  increase towards longer orbital periods. Therefore the analysis
  regarding the origin of the high WD masses in CVs has to be
  revisited, in particular in the context of the information held by
  the WD mass distribution of CV progenitors.

\section{From PCEBs to pre-CVs}

A large sample of PCEBs has been identified by follow-up observations
of WDMS binaries identified with SDSS. Here, we complement these SDSS
PCEBs with systems that have been identified independent of SDSS and
extract a sample of PCEBs that could be representative of the progenitors of
present day CVs.

\subsection{The PCEB sample}\label{sec:sample}

The sample of PCEBs analysed here consists of the 60 WDMS PCEBs with
accurate parameters (35 new systems identified by SDSS and 25
previously known systems) listed in \citet[][ hereafter
  Z10]{zorotovicetal10-1} with the exception that we excluded
SDSS$1648+2811$ because its WD mass is extremely uncertain ($\Mwd =
0.63 \pm 0.52$). We also added three PCEBs from SDSS for which we
determined orbital periods since the publication of Z10 (a full
overview on the PCEBs from SDSS for which orbital periods have been
measured will be presented in Nebot G\'omez-Mor\'an et al. 2011,
submitted). Hence, our final sample contains 62 PCEBs
(Table\,\ref{tab:prop}). Note that some of the masses listed in
Table\,\ref{tab:prop} slightly differ from those in Z10 because we
refined our spectral-fitting routine
\citep[see][]{rebassa-mansergasetal11-1}. The sixth column labelled
``Type'' indicates the most probable type of WD (He-core or C/O-core)
according to the mass derived from observations. For systems with
$\Mwd\sim 0.48-0.51$\,\Msun\ we cannot decide which type of WD is the
most probable because this range of WD masses is not allowed by stellar
evolution models\footnote{Strictly speaking, these primary masses may result from CE evolution
occurring when the core of the giant is not yet fully degenerate, which
leads to hybrid He$+$C/O core sdB post-CE primary stars.
However, here we ignore this channel because it is not an important
CV formation channel.}.
We determine the type of the WD (He-core or
C/O-core) for a given value of the CE efficiency $\alpha$ using our
reconstruction algorithm in Sect\,\ref{sec:res}.

\addtocounter{table}{2}

\subsection{The past and future evolution of known PCEBs}

The first step in our reconstruction is to compute the
time that passed since the binary left the CE ($t_{\mathrm{cool}}$), which is estimated by interpolating cooling tracks of
\citet[][]{althaus+benvenuto97-1} for He-core WDs and \citet[][]{wood95-1}
for C/O-core WDs. To reconstruct the post-CE evolution we follow
\citet{schreiber+gaensicke03-1}. As evidence for a discontinuity in
the braking law for low-mass secondary stars is growing 
\citep[][ and references therein]{schreiberetal10-1}, we
here assume the latest version of disrupted magnetic braking from
\citet{hurleyetal02-1} with the normalization provided by
\citet{davisetal08-1} to obtain the period the binary had
just after the CE phase ($P_{\mathrm{CE}}$), i.e.,
\small
\begin{equation}\label{eq:pce}
P_{\mathrm{CE}}=\left(\frac{3Ct_{\mathrm{cool}}(\Mwd + M_{\mathrm{2}})^{\frac{1}{3}}M_{\mathrm{2,e}}R_{\mathrm{2}}^3(2\pi)^{\frac{10}{3}}}{G^{\frac{2}{3}}\Mwd M_{\mathrm{2}}^2}+\Porb^{\frac{10}{3}}\right)^{\frac{3}{10}},
\end{equation}
\normalsize
where the masses and the radius of the secondary are in solar units,
the observed period (\Porb) is in years, $C=3.692\times10^{-16}$, and
$M_{\mathrm{2,e}}$ is the mass of the secondary's convective envelope,
given by
\small
\begin{equation}
M_{\mathrm{2,e}}=0.35\left(\frac{1.25-M_{\mathrm{2}}}{0.9}\right)^2,
\end{equation}
\normalsize
for $0.35\Msun \leq M_{\mathrm{2}} \leq 1.25\Msun$
\citep[see][]{hurleyetal00-1}. For $M_{\mathrm{2}} < 0.35\Msun$
we assume that the binary is losing angular momentum only due to
gravitational radiation.

We then use Eq.\,(11) from \citet[][ corrected as in
  Z10]{schreiber+gaensicke03-1} to compute the period the PCEBs will
have when becoming semi-detached again ($P_{\mathrm{sd}}$), i.e. the
period when they will start mass transfer as CVs.  Replacing
$P_{\mathrm{CE}}$ by $P_{\mathrm{sd}}$ and $t_{\mathrm{cool}}$ by
$-t_{\mathrm{sd}}$ in Eq.\,\ref{eq:pce} finally gives the corresponding
duration ($t_{\mathrm{sd}}$).

To compute the duration of the pre-CE evolution, i.e. the time the
systems needed to enter the first phase of mass transfer, we need to
derive the mass of the progenitor of the WD for each system. To
reconstruct the CE evolution of the PCEBs in our sample we use the
energy equation, including a fraction \ali\ of internal energy of the
envelope, and assume $\alpha = \ali = 0.25$ based on the results of Z10.
As in Z10 we include the effects of internal energy in the structural
parameter $\lambda$. The energy equation then becomes
\small
\begin{equation}\label{eq:alpha}
0.25\times\Delta E_{\mathrm{orb}} = −\frac{G M_{\mathrm{1}} M_{\mathrm{1,e}}}{\lambda_{\mathrm{0.25}} R_{\mathrm{1}}}
\end{equation}
\normalsize
where $M_{\mathrm{1}}$, $M_{\mathrm{1,e}}$ and $R_{\mathrm{1}}$ are
the total mass, envelope mass, and radius of the WD progenitor at
the onset of the mass transfer, and $\lambda_{\mathrm{0.25}}$ is the
structural parameter including a fraction $\ali = 0.25$ of the
internal energy available in the giants envelope. We further assume
that the observed WD mass is equal to the core mass of the giant
progenitor at the onset of mass transfer and use the single-star
evolution (SSE) code from \citet{hurleyetal00-1} to calculate
$R_{\mathrm{1}}$ as well as the equations from the latest version of
the binary-star evolution (BSE) code from
\citet{hurleyetal02-1}\footnote{The latest version of the BSE code provides an 
algorithm that computes $\lambda$ including a user-defined fraction of 
internal energy. This routine is based on fits to detailed stellar models 
from \citet{polsetal98-1}.}
to compute the values of $\lambda_{\mathrm{0.25}}$, for different values
of $M_{\mathrm{1}}$ (in steps of 0.01 \Msun). Using this in
Eq.\,\ref{eq:alpha}, we numerically obtain a solution for
$M_{\mathrm{1}}$. The SSE code also gives us for each system in our
sample the age ($t_{\mathrm{evol}}$) and the actual mass of the
progenitor (considering mass-loss) when it fills the Roche lobe. We
adopt a minimum initial mass of $0.96$\,\Msun\ for the progenitor
because stars with smaller initial masses have not had enough time to
evolve off the MS in less than $13.5 \times 10^{9}$\,yr. For more
details and a discussion of the assumption $\alpha = \ali = 0.25$ see
Z10.

Finally, the total time since the binary was born until it becomes a
CV, $t_{{\rm tot}}$, is given by the sum of the estimated lifetime of
the primary before it fills the Roche lobe, $t_{{\rm evol}}$, the
present cooling age, $t_{{\rm cool}}$, and the time the PCEB still
needs to become a CV, $t_{{\rm sd}}$.

\subsection{Pre-CVs: representative for progenitors of current 
CVs}\label{sec:repres}
Because the following analysis of CV progenitors is motivated by the
WD-mass distribution of the \textit{present-day} CV population,
we define a pre-CV to be a PCEB that could have been a
progenitor of a present-day CV. These systems need to have CV
formation times ($t_{{\rm tot}}$) shorter than the age of the Galaxy and
the second mass-transfer (CV) phase must be stable.

Assuming full conservation of mass and orbital angular momentum of the binary,
\citet{hjellming+webbink87-1} obtained a limit on the mass ratio for
dynamically-stable mass transfer of $M_{\mathrm{2}}/M_{\mathrm{1}} <
q_{\mathrm{crit}} = 0.634$ for complete polytropes 
that represent an excellent approximation for fully convective stars. 
According to \citet{warner95-1} this limit remains very similar,
$q_{\mathrm{crit}}\sim 2/3$, for secondaries up to 
$M_{\mathrm{2}} \leq 0.7-0.8$\,\Msun\ \citep[see also][ their
  Fig.\,2]{dekool92-1}.
Throughout this work we use $q_{\mathrm{crit}}=2/3$ 
(but note that for non-conservative mass transfer the value of
$q_{\mathrm{crit}}$ can significantly differ from this value). 
Systems with mass ratios exceeding $q_{\mathrm{crit}}$
and $M_{\mathrm{2}} \leq 0.7-0.8$\,\Msun\ experience 
unstable mass transfer that is likely to cause a second CE 
phase resulting in a merger or the generation of double-degenerate 
systems instead of CVs.
For more massive secondaries
($M_{\mathrm{2}}\,\gappr\,0.8\Msun$) the second mass-transfer phase can be
dynamically stable but thermally unstable if the mass ratio is
$q\,\gappr\,1$. These systems may evolve into CVs following a period of
thermal-timescale mass transfer \citep{rappaportetal94-2,
  schenker+king02-1, schenkeretal02-1}. However, because all the systems
in our sample have $M_2 < 0.6\Msun$ (except IK\,Peg which, owing to its
long orbital period and high secondary mass, will start mass transfer
when the secondary becomes a giant, and the system will probably
enter a second CE phase), we decided to use $q \leq 2/3$ as a limit
to become a CV for all the PCEBs in our sample.

For the age of the Galaxy we assume $13.5\times10^{9}$\,yr
\citep{pasquinietal04-1}.

\subsection{The future zero-age CV population}\label{sec:res}

Using the tools described in the previous section, we have
reconstructed the evolutionary history of the PCEBs in our sample as
well as calculated their future evolution into
CVs. Table\,\ref{tab:res} lists the corresponding results for each
system, i.e.  the mass of the WD we used for all the subsequent
computations (\Mwd), the current mass ratio ($q =M_\mathrm{2}/\Mwd$),
the period the PCEB had just after the CE phase ($P_{\mathrm{CE}}$),
the initial mass of the progenitor of the WD ($M_\mathrm{1,i}$), the
mass of the progenitor of the WD when it fills the Roche lobe
($M_\mathrm{1,CE}$), the initial orbital separation
($a_{\mathrm{i}}$), the period that it will have when it becomes a CV
($P_{\mathrm{sd}}$), and the durations of the different evolutionary
phases as well as the total CV formation time ($t_{\mathrm{evolv}}$,
$t_{\mathrm{cool}}$, $t_{\mathrm{sd}}$, $t_{\mathrm{tot}}$). Pre-CVs,
i.e. PCEBs that are representative of the progenitors of the
present-day CV population, are set in bold. Note that $\Mwd$ slightly
differs in some cases from the spectroscopic mass estimate because, e.g., WD
masses in the range of $\Mwd\sim 0.48-0.51$\,\Msun\ contradict
fundamental predictions of CE evolution. As noticed, e.g., by
\citet[][]{politano96-1} WD masses in this range should not appear in
the mass distribution of WDs in PCEBs because they are larger than the
maximum core mass of stars on the FGB and smaller than the core mass
on the AGB having the same radius.
In these cases we use the WD mass closest to the observed value in
Table\,\ref{tab:prop} that provides a solution for $\alpha = \ali$
in the range of $0.2-0.3$. Because the real value of $\alpha$ is
still uncertain, we give priority to find a core mass closer to the
observed WD mass for these systems, allowing $\alpha$
to vary within this range. Although the fraction of systems with
He-core WDs may be slightly different for different values of $\alpha$,
the main conclusions of this work are not affected by this uncertainty.

\addtocounter{table}{3}

\subsubsection{Notes on individual systems} 

For five systems containing He-core WDs (EC$13349-3237$, UX\,CVn, SDSS$1724+5620$,
SDSS$1731+6233$, and SDSS$2123+0024$), we could not determine the cooling age because
the cooling tracks from \citet{althaus+benvenuto97-1} corresponding to
their masses start at lower temperatures. However, this means that
these systems are young enough ($t_{\mathrm{cool}} < 0.05$\,Gyr) to
neglect the cooling age and to assume the current period to be very
similar to the period the PCEB had at the end of the CE phase,
i.e. $P_{\mathrm{CE}} \simeq P_{\mathrm{obs}}$.

SDSS$0052-0053$ has notionally $t_{\mathrm{sd}} < 0$, i.e. according
to the mass ratio and measured period, this systems should already be
a CV.  We exclude this system from our list of pre-CVs because the
small separation implies that it might not be a PCEB, but a detached
CV in the period gap \citep{davisetal08-1}.

WD$0137-3457$ is a well-studied PCEB containing a WD plus brown dwarf
\citep{burleighetal06-1}, which will start mass transfer at a very
short period below the orbital-period minimum. 

\subsubsection{Mass-ratio distribution}

\begin{figure}
\centering
\includegraphics[angle=270,width=0.49\textwidth]{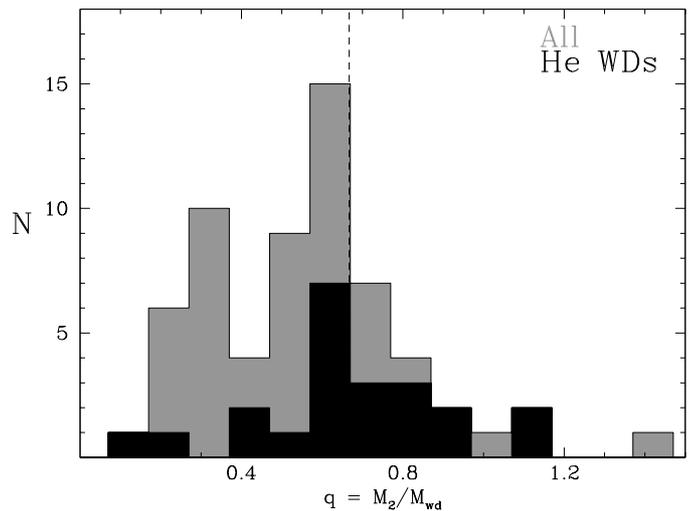}
\caption[]{Mass-ratio distribution for the 62 PCEBs in our sample. Black is for systems with He-core WDs and grey is for all the systems. The dashed line indicates the limit for stable mass transfer.}
\label{fig:q}
\end{figure}

Inspecting Table.\,\ref{tab:res}, our sample contains 22 PCEBs with He-core ($\Mwd < 0.5$) and 40 with C/O-core WDs ($\Mwd > 0.5$\,\Msun), corresponding to a He-core WD fraction of $35 \pm 6\%$. Figure\,\ref{fig:q} shows the mass-ratio distribution of the 62 PCEBs. Applying the criterion for $q$ outlined in Sect.\,\ref{sec:repres}, only 43 of the 62 systems will evolve into CVs with stable mass transfer. Among those, 11 ($26 \pm 7\%$) contain a He-core WD (black histogram in Fig.\,\ref{fig:q}).

\subsubsection{Evolution timescales}

\begin{figure}
\centering
\includegraphics[width=0.49\textwidth]{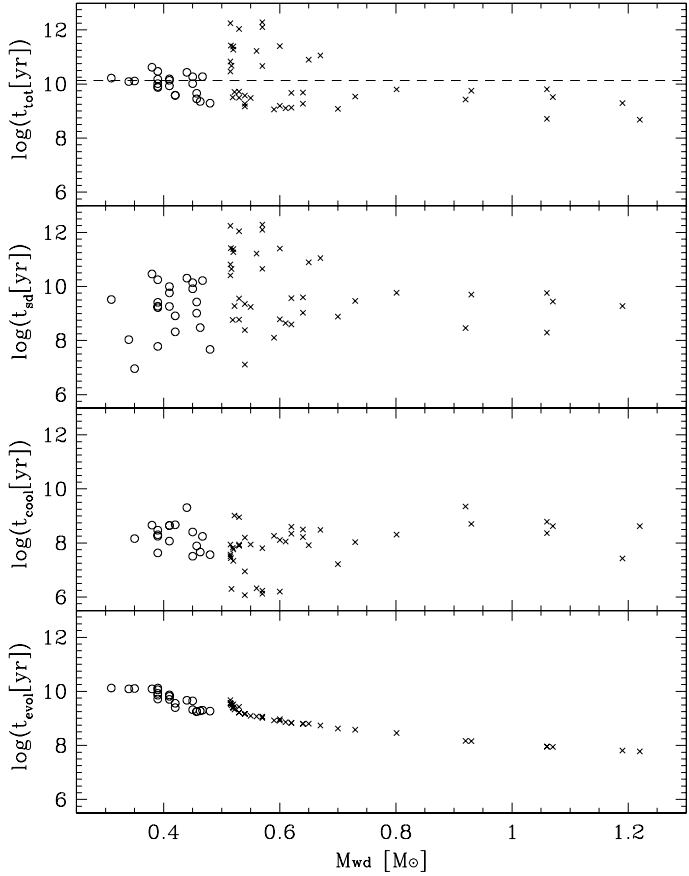}
\caption[]{Different timescales computed for the 62 PCEBs in our sample. Open circles are for systems with He-core WDs and crosses for systems with C/O-core WDs. \textit{From top to bottom}: total time since the systems were born until they become CVs; time the actual PCEBs need to become CVs; cooling time; and time since the systems were born until the CE phase. The dashed line in the top panel correspond to $13.5 \times 10^{9}$\,yr.}
\label{fig:times}
\end{figure}

Figure\,\ref{fig:times} shows the different timescales we computed for all systems. We use different symbols to distinguish between systems with He-core and C/O-core WDs. The main difference between the two populations is related to the pre-CE evolution time ($t_{\mathrm{evol}}$), which is continuously decreasing with increasing \Mwd. The PCEBs containing He-core WDs need at least $10^9$\,yr to be formed \citep[see also][]{politano96-1}, i.e. these systems were already old when they entered the CE phase. In contrast, PCEBs containing C/O-core WDs can have massive progenitors and short pre-CE lifetimes. It is also evident from Fig.\,\ref{fig:times} that, in general, our sample is dominated by systems with $t_{\mathrm{cool}}<<t_{\mathrm{sd}}$, which implies that most currently observed PCEBs have so far only completed a small fraction of the time they will spend as PCEBs. This may introduce a minor bias against high-mass WDs in pre-CVs because the undetected systems with longer $t_{\mathrm{cool}}$ need to have shorter $t_{\mathrm{evol}}$ to become CVs within the age of the galaxy.

If we apply both cuts in $q$ and $t_{\mathrm{tot}}$, we are left with 24 systems that we consider to be genuine pre-CVs, i.e. representative for the progenitors of the current-day CV population. The fraction among these pre-CVs that have He-core WDs is $17 \pm 8\%$ (4 of 24 systems).

\subsubsection{The WD-mass and orbital-period distribution}

\begin{figure}
\centering
\includegraphics[angle=270,width=0.49\textwidth]{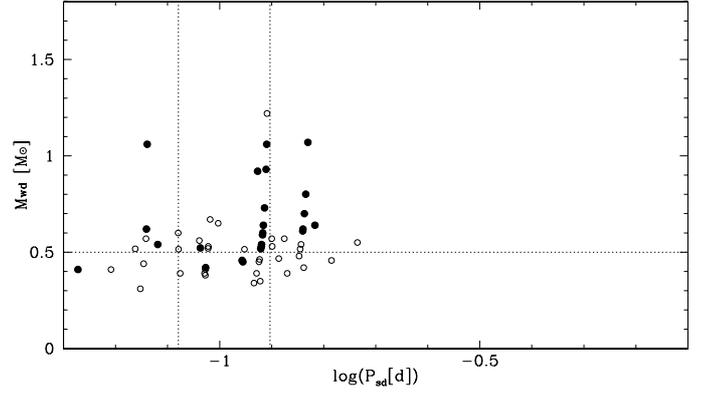}
\caption{PCEB WD masses as a function of the period they will have when evolving into semi-detached CVs. Systems that are genuine pre-CVs (according to their mass ratios and total CV-formation times) are shown as solid symbols. The dotted lines are the same as in Fig.\,\ref{fig:cvmass}.
}
\label{fig:PCEBmass}
\end{figure}

Figure\,\ref{fig:PCEBmass} shows the relation between the mass of the WDs in our sample and the orbital period they will have when they start the future CV mass-transfer phase. Most PCEBs are going to start the mass transfer above or in the period gap. Pre-CVs (solid symbols) containing low-mass WDs ($< 0.5$\,\Msun) tend to start the mass transfer in or below the gap. This is consistent with the results of \citet{dekool92-1} and \citet{politano96-1}, who predicted that CVs that start mass transfer in or below the gap are usually older than those formed above the gap, and that the CV population above the gap should not contain systems with He-core WDs.

The average mass of the pre-CV WDs is $0.67\pm0.20\,\Msun$,
which is slightly larger than the average for single WDs, but still well below the observed average for CVs.

\subsection{Potential selection effects in the PCEB sample}\label{sec:pcebbias}

In Sect.\ref{sec:cvbias} we have shown that the WD-mass distribution of CVs from SDSS is not significantly biased. Before comparing the pre-CV WD-mass distribution obtained here with the WD-mass distribution of CVs from Sect.~\ref{sec:cvmass}, we have to carefully inspect potential observational selection effects for our PCEB and pre-CV samples that could have affected our pre-CV WD-mass distribution.

\subsubsection{Identifying PCEBs in SDSS}

Identifying WDMS binaries using SDSS spectra requires both stellar components to be clearly visible in the spectrum. This introduces a bias towards bigger less massive WDs that are easier to detect against a companion. This effect has recently been analysed in detail by  \citet{rebassa-mansergasetal11-1}, who obtained the following result. Assuming a scale height of $200$\,pc and considering only systems containing WDs hotter than $12\,000$\,K and M-dwarf companions, the detection of a PCEB containing a He-core $0.4\Msun$ WD is $\sim1.5$ times more likely than of a PCEB containing a C/O-core $0.6\Msun$ WD. This implies that the PCEB and pre-CV samples analysed here are also slightly biased towards systems containing low-mass WDs and that the intrinsic fraction of He-core pre-CVs is likely lower than the $17\pm8\%$ we obtained.

\subsubsection{Bias against early-type companions}\label{subsec:etbias}

\begin{figure}
\centering
\includegraphics[width=0.49\textwidth]{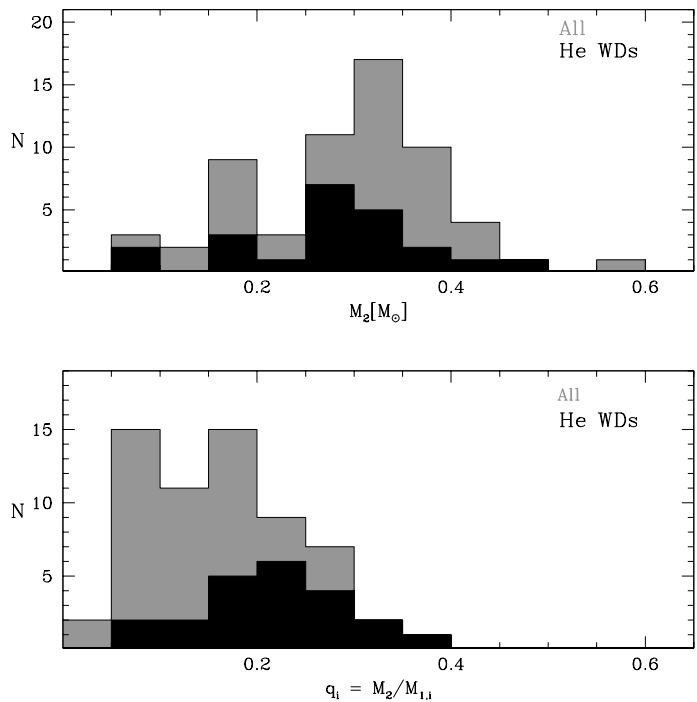}
\caption[]{\textit{Top}: Secondary-mass distribution of the PCEBs in our sample.  \textit{Bottom}: Initial mass-ratio distribution. Apparently, the mass distribution peaks at $\sim 0.3-0.35$\,\Msun, close to the limit for a He-core WD companion with stable mass transfer. The initial mass ratio increases towards low values with a steep decline for very low values. Black is for systems containing He-core WDs, while grey is for the whole sample. (IK\,Peg was excluded from the top panel owing to its high secondary mass compared to the rest of the sample)}
\label{fig:m2}
\end{figure}

Stable mass transfer (in the second mass-transfer phase of the binary) is one of the crucial conditions for a PCEB to become a CV, and therefore possible biases concerning the mass of the secondary star also have to be considered. The PCEBs with secondary stars earlier than $\simeq$ M3 are underrepresented in our sample because of two reasons. First, because magnetic braking is inefficient in fully convective stars but a strong source of angular momentum loss in systems with non-fully convective secondary stars ($M_2\,\gappr\,0.35$\Msun), the latter spend significantly less time in the evolutionary phase between the CE and the second mass-transfer phase \citep{politano+weiler06-1,schreiberetal10-1}. Second, our PCEB survey \citep[e.g.][]{rebassa-mansergasetal07-1} is limited to M-dwarf secondary stars and we therefore entirely miss pre-CVs with K- to F-type companions.  The mass distribution of the secondaries in our sample, shown in the top panel of Fig.\,\ref{fig:m2}, peaks at $\sim 0.3-0.35$\,\Msun\ (spectral type M3--M4) and steeply declines towards higher masses. Because PCEBs with secondaries earlier than M3 must contain C/O-core WDs to become stable-mass-transfer systems, the bias against high-mass secondary stars in PCEBs implies a bias against high-mass WDs in pre-CVs. Whether or not this bias results in a dramatic bias against high-mass WDs in the emerging pre-CV sample depends entirely on the initial-mass-ratio distribution.

In the bottom panel of Fig.\,\ref{fig:m2} we show the initial-mass-ratio distribution reconstructed from our systems. The shape is similar to those obtained in most studies of the mass-ratio distributions of MS binaries in the range of $0\leq\,q\leq\,0.5$, i.e. the distribution increases towards low values of $q=M_{\mathrm{2}}/M_{\mathrm{1,i}}$ and steeply declines for very low values $q\,\lappr\,0.1$ \citep[see e.g.][]{trimble90-1,duquennoy91-1,mazehetal98-1}. This seems to support the assumption of an initial-mass-ratio distribution favouring unequal masses. One should, however, keep in mind that the distribution shown in Fig.\,\ref{fig:m2} may significantly differ from true the initial-mass-ratio distribution because of two reasons. First, the initial-mass-ratio distribution reconstructed from PCEBs might be slightly different from the intrinsic mass-ratio distribution because a higher fraction of systems containing very low-mass secondaries may merge instead of expelling the envelope. Second, and probably more important, as we already mentioned our PCEB sample is biased against systems containing secondary stars earlier than M3, which might have strongly contributed to the observed decline in the number of systems towards larger mass ratios.

\subsubsection{Spectroscopic WD masses}

As discussed in detail in \citet{rebassa-mansergasetal10-1}, recent work \citep{falconetal10-1,tremblay+bergeron09-1} suggests that WD mass determinations from fitting model atmospheres to observed spectra may underestimate the masses of the WDs. Because the WD masses used in this work have been obtained with such an algorithm combined with spectral-decomposition methods \citep{rebassa-mansergasetal10-1}, it might be that the WD masses given here are systematically smaller than the true WD masses by $0.03-0.05$\Msun. This may again reduce the intrinsic number of PCEBs and pre-CVs containing He-core WDs but will certainly not change the overall appearance of the WD-mass distribution.\\

In summary, all the possible biases mentioned above point towards a higher intrinsic mean mass of the WDs in our PCEB sample. How dramatic these biases are mainly depends on the still unknown initial-mass-ratio distribution, as we will discuss in more detail in Sect.\,\ref{sec:disc}

\section{Comparing pre-CV, PCEB, and CV WD-mass distributions}\label{sec:compare}

\begin{figure}
\centering
\includegraphics[width=0.49\textwidth]{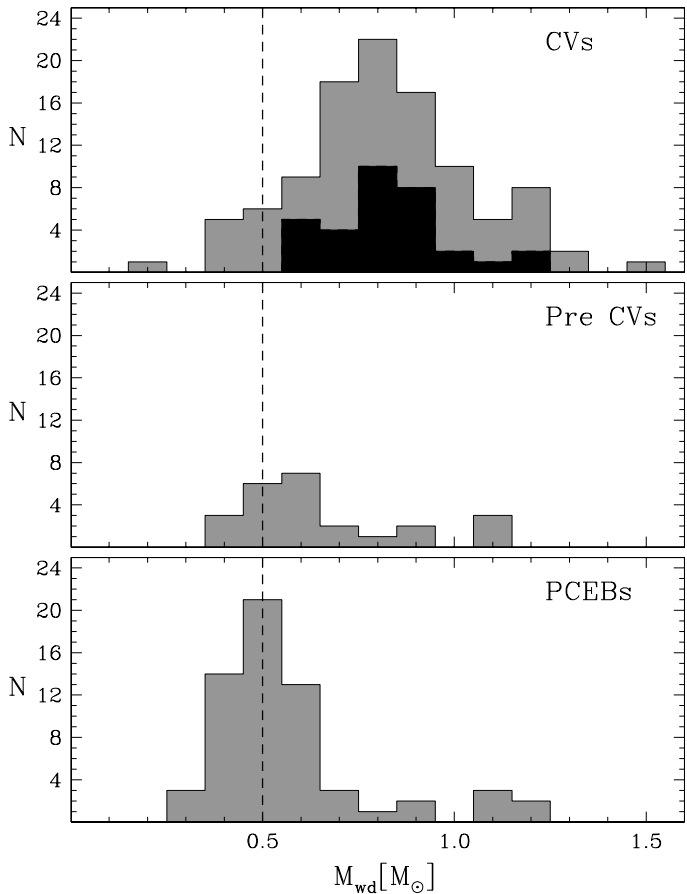}
\caption[]{Mass distribution of the WDs in CVs (\textit{top}), pre-CVs (\textit{middle}), and PCEBs (\textit{bottom}). The black histogram in the top panel represents the 32 fiducial CV WDs with presumably more reliable mass, defined in Sect.\,\ref{sec:cvmass}.}
\label{fig:mwdall}
\end{figure}

After reviewing the measurements of the WD masses in CVs and determined the WD-mass distributions of PCEBs and pre-CVs, we may now compare these distributions. Figure\,\ref{fig:mwdall} shows the WD-mass distributions of CVs (top), pre-CVs (middle), and PCEBs (bottom). The main difference between PCEBs and pre-CVs is that in the latter we had to exclude mostly systems with low-mass WDs, mainly because of their long evolutionary timescales or unstable mass ratios, an exclusion consistent with the lack of low-mass WDs in CVs. The mean WD mass for PCEBs is $\langle\Mwd\rangle=0.58\pm0.20$\,\Msun\, and it increases to $\langle\Mwd\rangle=0.67\pm0.20$\,\Msun\, for pre-CVs. The low-mass tail of the CV and pre-CV distributions look similar, i.e. increasing towards higher masses up to $\sim 0.6$\,\Msun.  However, a dramatic difference between the distributions is that CVs are dominated by systems containing high-mass WDs ($\Mwd\sim0.8\Msun$), which are nearly absent in the pre-CV or PCEB sample.

\begin{figure}
\centering
\includegraphics[width=0.49\textwidth]{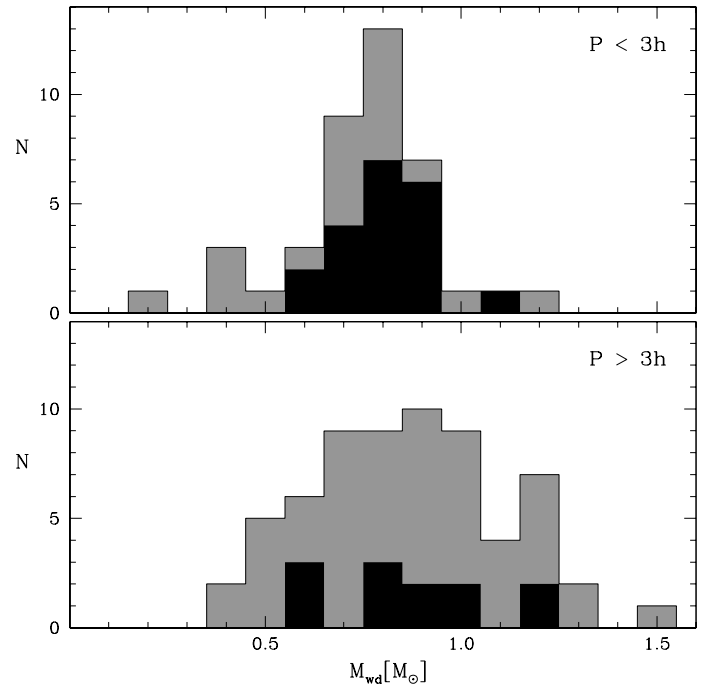} 
\caption[]{CV WD-mass distribution divided into systems in or below the period gap (\textit{top}), and above (\textit{bottom}). As in Fig.\,\ref{fig:mwdall}, black histograms correspond to the 32 fiducial CV WDs.}
\label{fig:mwdcvp}
\end{figure}

Important information concerning the origin of the high WD masses in CVs might be provided by inspecting the dependence on the orbital period. In Fig.\,\ref{fig:mwdcvp} we divided the WD-mass distribution of CVs into systems that are in or below the period gap (top panel) and systems above the period gap (bottom). As mentioned in Sect.\,\ref{sec:cvmass}, the mean WD masses do not differ significantly, however, systems above the period gap show a broad distribution, while systems with $\Porb < 3$\,h show a strong peak at $\sim 0.8$\,\Msun. This result remains if we only consider those systems with the most reliable WD mass determinations, the fiducial sub-sample of 32 systems (black histogram in both panels).

However, we advocate some caution when interpreting this result: The fiducial sub-sample is subject to small number statistics and the apparent difference between short- and long-orbital-period systems might be related to the more inhomogeneous combination of methods used for WD mass determinations in CVs above the gap (see Table\,\ref{tab:cvmass}), which might introduce significant scatter.

\section{Discussion}\label{sec:disc}

The main results obtained in the previous sections can be summarized as follows. The masses of CV WDs do significantly exceed the WD masses of both pre-CVs and single WDs. The high masses of CV WDs cannot be explained as the consequence of observational biases. The mean CV WD masses below and above the period gap are very similar but the mass distribution seems to be broader above the gap. Finally, very few, if any, He-core WD candidates are known in CVs and the fraction of He-core WDs in pre-CVs is $\lappr\,17\pm8\%$. Below we briefly review the predictions of binary-population-synthesis (BPS) models of CVs before discussing possible explanations for the high mean WD masses in CVs.

\subsection{BPS models of CVs}\label{subsec:comp}

The first BPS model of CVs and their progenitors has been presented in a pioneering work by \citet{dekool92-1}, who calculated the formation rates of CVs for different initial-mass-ratio distributions and CE efficiencies \citep[for similar studies see][]{dekool+ritter93-1,kolb93-1,kingetal94-1,politano96-1,howelletal01-1}. The conclusions of these early works can be summarized as follows.

\begin{itemize}

\item The most crucial uncertain parameters are the initial-mass-ratio distribution and the CE efficiency. The peak of the orbital-period distribution predicted by BPS models of PCEBs depends crucially on the initial-mass-ratio distribution, while the width of the peak depends on the CE efficiency $\alpha$.

\item The WD-mass distribution of newly formed CVs is bimodal with maxima at $\sim0.4\Msun$ (He-core WDs) and $\lappr\,0.6\Msun$ (C/O-core WDs).  The fraction of He-core WDs in CVs can be up to $\sim50\%$ but decreases for low values of the CE efficiency.

\item If the initial-mass-ratio distribution favours unequal masses, most CVs are born below the gap.

\item If the initial-mass-ratio distribution favours equal masses and/or the CE efficiency is small and/or CVs descending from super soft X-ray binaries are included, most CVs might be born above the orbital-period gap.

\item Most importantly in the context of this paper, the predicted mean WD mass of BPS models for CVs are typically well below the mean WD mass of single stars \citep[see e.g.][who predicts $\langle\Mwd\rangle=0.49\Msun$]{politano96-1}.

\end{itemize}

\subsection{Initial-to-final mass relations}\label{sec:ifmr}

\begin{figure*}
\centering
\includegraphics[angle=270,width=0.75\textwidth]{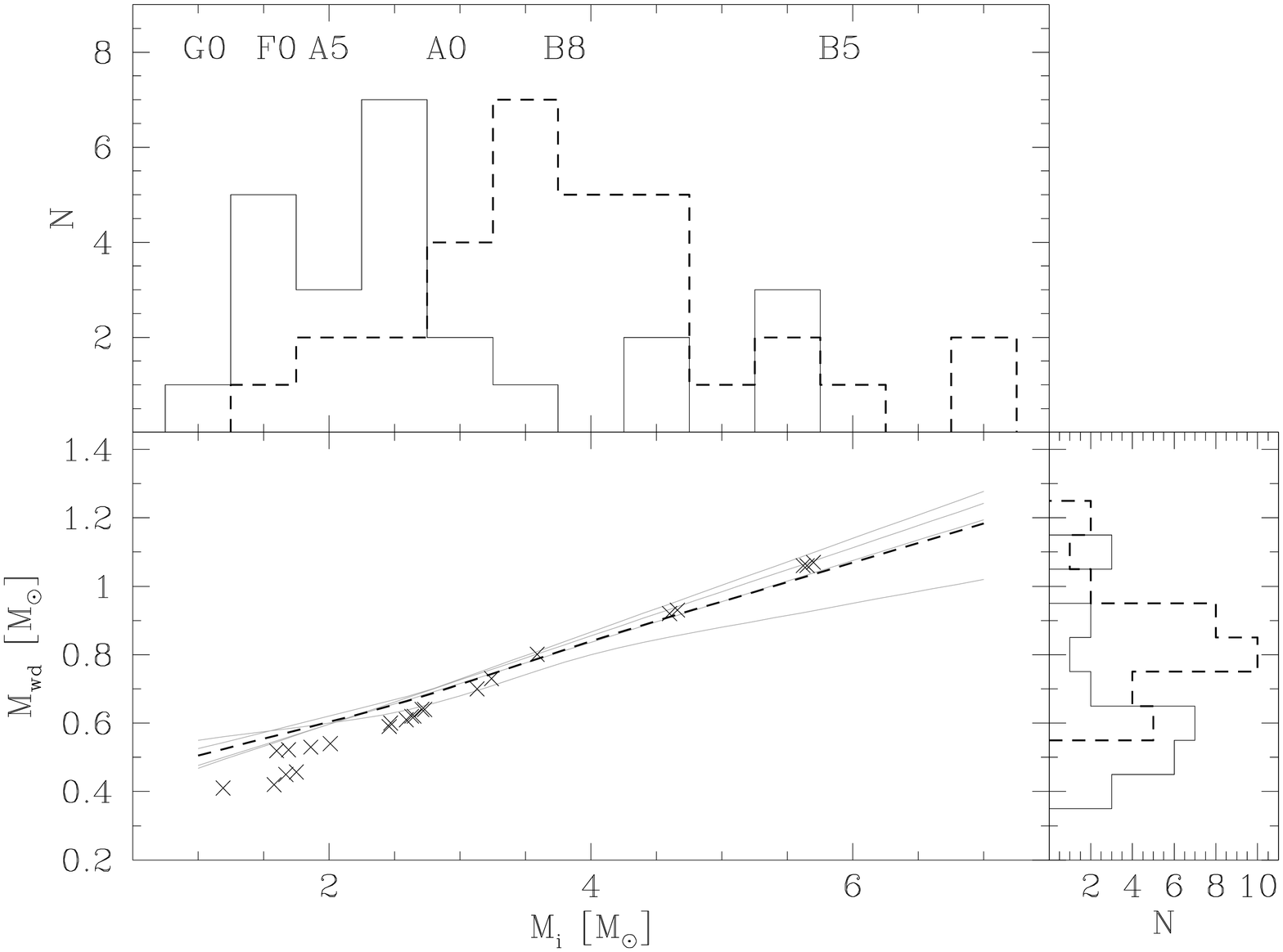}
\caption{\textit{Right:} Distribution of WD masses for the pre-CVs (solid) and for the fiducial sample of CVs (dashed). \textit{Bottom left:}  Comparison between the initial-to-final mass (IFM) of our sample of pre-CVs (crosses) with several proposed IFM relations for single-star evolution (grey lines, from top to bottom: \citealt{catalanetal08-1}, \citealt{williamsetal09-1}, \citealt{casewelletal09-1}, and \citealt{weidemannetal00-1}) and their average (black dashed line). For WD masses exceeding $\sim\,0.5\Msun$ the reconstructed initial masses are very close to the average IFM.\textit{Top:} Progenitor-mass distribution for the WDs in pre-CVs (solid) and for the WDs in CVs (dashed). The latter was computed assuming the average IFM relation from the bottom left panel, which gives reasonable results because of the high WD masses in our fiducial CV sample (see the text for details). }
\label{fig:IFM}
\end{figure*}

Before discussing the implications of our results for the standard theory of CV evolution, we emphasize and illustrate our main finding once more, i.e. the obtained difference of WD masses in pre-CVs and CVs. The right-hand side of Fig.\,\ref{fig:IFM} shows the distribution of WD masses for the pre-CVs in our sample (solid histogram) and for the sample of 32 fiducial CVs (dashed histogram). In the bottom left panel we compare the initial-to-final mass (IFM) of the WDs in our pre-CV sample (crosses) with different theoretical and empirical IFM relations for single-star evolution (grey lines, from top to bottom: \citealt{catalanetal08-1}, \citealt{williamsetal09-1}, \citealt{casewelletal09-1}, and \citealt{weidemannetal00-1}). The black dashed line corresponds to the average of the four IFM relations. The differences between the reconstructed initial masses for the pre-CVs and the average IFM relation are negligible for $\Mwd\,\gappr\,0.5\Msun$, which is true for all fiducial CVs. Therefore we can use the average IFM relation to obtain a reasonable guess of the initial mass for the WDs in CVs. The top panel, finally, compares the distribution of reconstructed initial masses for the WDs in our sample of pre-CVs (solid histogram) with the distribution of initial masses for the WDs in the 32 fiducial CVs obtained from the average IFM relation (dashed histogram). Even considering that there might be an error of $\sim\,0.5\Msun$ in the progenitor masses for the CV WDs, the far-reaching consequences of the high WD masses in CVs are evident: Either CV WDs grow in mass, or the vast majority of the currently known CV WDs must be the descendants of very massive progenitors ($M_i\,\gappr\,3-4$\Msun). In the following sections we discuss both options.

\subsection{Mass growth in CVs?} \label{sec:growth}

First, we assume that the initial-mass-ratio distribution favours unequal masses and that both the measured mean WD masses of CVs and pre-CVs are good approximations of the corresponding intrinsic population. In this case, CV WD masses must grow through accretion as suggested also by \citet{savouryetal11-1}. Most CVs below or in the gap would have been born there. The formation time would be dominated by the slow post-CE evolution driven only by angular momentum loss through gravitational radiation with total CV formation times longer than $10^9$\,yr. Because virtually all WD masses are allowed by the stability criterion, one would expect a superposition of the corresponding pre-CV WD-mass distributions shifted towards higher masses depending on their age as a CV. Furthermore, one would expect a slight increase of the WD mass towards shorter orbital periods below the gap. Inspecting Figs.\,\ref{fig:cvmass} and \ref{fig:mwdcvp} one might like to speculate that this is the case but more high-quality WD-mass measurements of CVs are certainly required. In contrast, CV formation above the gap would have started earlier because magnetic braking significantly reduces the time between the CE and CV phases. For each starting orbital period a different range of WD masses is allowed and the evolution towards shorter orbital periods is fast. One would therefore expect a fairly broad distribution similar to the observed one.

However, this scenario has the obvious and dramatic disadvantage that it contradicts all current theories of nova eruptions. Simulations of nova outbursts predict that the amount of ejected mass is larger than the amount of accreted matter \citep[see e.g.][]{prialnik+kovetz95-1,yaronetal05-1}, except in the case of very high mass-transfer rates ($\gappr\,10^{-8} \Msun\,\mathrm{yr}^{-1}$), which implies that in most of the cases the mass of the WD decreases during the CV phase. The observed abundances of nova ejecta are found to be enriched in heavy elements, which is used as an argument in favour of core erosion \citep[see e.g.][]{truran+livio86-1,livio+truran94-1}. If mixing of accreted material with the core of the WD is not very efficient, the observed abundances can only be explained by expelling more mass than has been accreted \citep{livio+truran92-1}. However, the mixing theory is far from being understood, and there is not even agreement on the most important mechanisms. In the last decade, some efficient mechanisms have been proposed, such as gravitational waves at the interface between the envelope and the WD \citep{alexakisetal04-1}, or Kelvin-Helmholtz instabilities that occur when the interface becomes convectively unstable \citep{casanovaetal10-1}. The amount of enhancement these studies obtain seems to agree with the observations. In addition, the amount of ejected mass predicted by the models is very sensitive on the nuclear-reaction rates used in the model calculations, and also on the specific nuclear reactions that are taken into account \citep[see e.g.][]{starrfieldetal09-1}.  Finally, one key point to keep in mind is that the observed classical novae are \textit{not at all} representative of the entire CV population, because the nova recurrence time is decreasing with increasing mass-transfer rate and increasing WD mass, and therefore the classical novae are strongly biased towards high-mass WDs in novalike variables with high mass-transfer rates.

In summary, observations of nova ejecta and theoretical predictions seem to indicate erosion of the WDs in CVs during a nova cycle, but considering the remaining caveats we tend to not generally exclude the possibility of WD mass growth in CVs.

\subsection{Are the known pre-CVs just the tip of the iceberg?}
\label{sec:sss}

The second option to explain the high WD masses in CVs considers the possibility that the currently known sample of pre-CVs presented here is not really representing the intrinsic properties of the galactic pre-CV population.  We already know that there are strong selection effects disfavouring companion stars with spectral types earlier than $\simeq$ M3, i.e. masses $\gappr\,0.35$\,\Msun, both because of the limited contrast ratio of the optical surveys (predominantly SDSS), and because of efficient magnetic braking in these systems that leads to short evolutionary timescales. This bias can be dramatic if the initial-mass-ratio distribution is flat, i.e. $dN\propto\,dq$, or even worse if it favours the formation of binaries with equal masses \citep[often referred to as the ``twin peak'', see e.g.,][]{soederhjelm07-1,lucy06-1,simon+obbie09-1}. In this case, the intrinsic population of pre-CVs would be dominated by systems with relatively massive secondary stars.  For pre-CVs with massive secondaries the mass-ratio limit for stable mass transfer requires the WD masses to be relatively high ($\Mwd>M_2/q_{\mathrm{crit}}$) in order to start stable mass transfer, or the binary must evolve through a period of thermal-timescale mass transfer before it becomes a CV. In the latter case the system appears as a super-soft X-ray source with stable nuclear burning on the surface of the WD allowing the WD mass to grow \citep{schenkeretal02-1}. Indeed, direct evidence for CVs descending from super-soft sources has been provided by \citet{gaensickeetal03-1}, who find CNO abundances to agree with previous thermal-timescale mass transfer in $\sim10-15\%$ of CVs with UV spectra. Hence, CVs descending from pre-CVs with high-mass secondaries, which have not been identified so far, may represent an important fraction of the current CV population. However, whether the bias against high-mass secondaries in our pre-CV sample can be strong enough to explain the high WD masses of the observed CV sample has to be clarified by performing BPS models. In addition, if indeed the vast majority of CVs below the gap are descendants from CVs above the gap, one would expect the shape of the CV WD-mass distributions above and below the gap to be very similar, which does not seem to be the case (see Fig.\ref{fig:mwdcvp} and the discussion in the corresponding section).

\subsection{A note on Type Ia supernovae progenitors}

The two currently most popular candidates for Type Ia supernovae (SNe Ia) progenitors are either close double C/O-core WDs with a total mass exceeding the Chandrasekhar limit (the double-degenerate channel, see e.g. \citealt{webbink84-1}; \citealt{iben+tutukov84-1}) or super-soft X-ray sources, i.e. close WDMS binaries with thermal-timescale mass transfer (the single-degenerate channel, see e.g. \citealt{vandenheuveletal92-1, kahabka+vandenheuvel97-1}). Recent studies of the time-delay distribution of SNe Ia (see e.g. \citealt{maoz10-1}, and references therein) indicate the existence of two populations of SNe Ia, a prompt channel that leads to explosions in $100-500$\,Myr and a delayed population that produces SNe Ia on much longer timescales \citep[$\sim5$Gyr,][]{mannuccietal06-1,scannapieco+bildsten05-1,sullivan06-1}.

Both options proposed above as possible explanation of the high WD masses in CVs may have implications for the still open question of which objects actually produce SNe Ia. If most CVs descend from super-soft sources, great numbers of SNe Ia emerging from the single-degenerate channel mostly with short delay times are expected.  If, on the other hand, CV WDs grow in mass through accretion, those born with sufficiently large WD masses may explode as a SNe Ia. Owing to their long formation times and their relatively low accretion rates, CVs would contribute to the long delay time SNe Ia population.

We note in passing that the importance of the single-degenerate channel is questioned by recent observational results.  The observed X-ray fluxes of early-type galaxies seems to be too low to be consistent with the prediction of large populations of super-soft sources \citep{gilfanov+bogdan10-1,distefano10-1} unless the super-soft phase is very short \citep{hachisuetal10-1}. Also, the non-detection of radio emission in a sample of 27 SNe Ia \citep{panagiaetal06-1} appears to contradict the single-degenerate scenario unless the accretion process onto the WD in the SN progenitor system proceeds without significant systemic mass loss to the interstellar medium.

\section{Conclusion} \label{sec:conc}

We have investigated the WD-mass distributions of CVs and pre-CVs and draw the following conclusions.

\begin{itemize}

\item The fraction of He-core WDs is small in both samples i.e. $\leq17 \pm 8\%$ (pre-CVs) and $\leq10\%$ (CVs), which agrees better with BPS models if a low efficiency for CE evolution ($\alpha \sim 0.25$) is assumed.

\item There is no significant evolution of WD mass with orbital period.

\item The mean WD mass of CVs significantly exceeds that of pre-CVs and PCEBs. This cannot be explained as an observational bias towards the detection of CVs containing high-mass WDs. 

\end{itemize}

To the best of our knowledge, the two only plausible options to explain the measured high mean WD masses in CVs are either a large population of PCEBs with early-type secondaries (partly evolving into CVs through thermal-timescale mass transfer) that has so far not been identified, or WD mass growth in CVs through accretion. While the latter option contradicts theories of nova eruptions, the first implies an initial-mass-ratio distribution that favours equal masses, which contradicts most observational studies of MS binaries. Both alternatives are very uncomfortable solutions, and proving either of them correct would have far-reaching implications for our understanding of compact-binary formation and evolution. Both alternatives also have possible implications for the pathways to SN\,Ia: the first option would imply a large population of super-soft X-ray binaries, representing an important fraction of CV progenitors; the second option would imply that a fraction of main-stream CVs can become SN\,Ia with long delay times. A crucial step towards a more definitive answer to this question is to overcome the observational biases against PCEBs with F, G, or K-type companions.

\begin{acknowledgements}
 MZ is supported by an ESO studentship. MRS acknowledges support from FONDECYT (grant 1061199 and 1100782).
We thank the referee, Ronald Webbink, for his excellent comments that helped to improve this paper.
\end{acknowledgements}

\bibliographystyle{aa}
\bibliography{aabib,aamnem99,aabib,proceedings}

\longtab{1}{
\begin{longtable}{lccccll}
\caption{\label{tab:cvmass} CVs with robust WD mass measurements, based on the analysis of eclipse light curves (e), radial velocity curves (d), gravitational redshifts (g), or spectrophotometric modelling (sp).}
\\
\hline\hline
\noalign{\smallskip}
System        & $\Porb$ & $\Mwd$ & $\sigma\Mwd$ & $M_2$ & Method & References\\
                & [min]   & [\Msun]& [\Msun] & [\Msun] & &  \& Notes\\
\noalign{\smallskip}
\hline
\endhead
\noalign{\smallskip}
\hline
\endfoot
\noalign{\smallskip}
SDSS$1507+5230$   &  66.6 & 0.892  & 0.008  & 0.058 & e      & 1,2,3 \\
SDSS$1433+1011$   &  78.1 & 0.865  & 0.005  & 0.057 & e      & 1,4,5\\
WZ\,Sge           &  81.6 & 0.85   & 0.04   & 0.078 & d,g,sp & 6,7 \\
SDSS$1501+5501$   &  81.9 & 0.767  & 0.027  & 0.077 & e      & 1,4 \\
SDSS$1035+0551$   &  82.1 & 0.835  & 0.009  & 0.048 & e      & 1,8 \\
SDSS$1502+3334$   &  84.8 & 0.709  & 0.004  & 0.078 & e      & 1,4 \\
SDSS$0903+3300$   &  85.1 & 0.872  & 0.011  & 0.099 & e      & 1,4 \\
XZ\,Eri           &  88.1 & 0.769  & 0.017  & 0.091 & e      & 1,9 \\
SDSS$1227+5139$   &  90.7 & 0.796  & 0.018  & 0.089 & e      & 1 \\
OY\,Car           &  90.9 & 0.84   & 0.04   & 0.086 & e      & 4,10 \\
CTCV$2354-4700$   &  94.4 & 0.935  & 0.031  & 0.103 & e      & 1 \\
SDSS$1152+4049$   &  97.5 & 0.560  & 0.028  & 0.087 & e      & 1 \\
OU\,Vir           & 104.7 & 0.703  & 0.012  & 0.115 & e      & 1,11,12 \\
HT\,Cas           & 106.1 & 0.61   & 0.04   & 0.09  & e      & 13 \\
IY\,UMa           & 106.4 & 0.79   & 0.04   & 0.10  & e      & 14 \\
VW\,Hyi           & 107.0 & 0.71   & 0.22   & 0.11  & g      & 15,16 \\
Z\,Cha            & 107.3 & 0.84   & 0.09   & 0.125 & e,d    & 17,18 \\
DV\,UMa           & 123.6 & 1.098  & 0.024  & 0.195 & e      & 1,9 \\
CTCV$1300-3052$   & 128.1 & 0.736  & 0.014  & 0.177 & e      & 1 \\
SDSS$1702+3229$   & 144.1 & 0.91   & 0.03   & 0.226 & e      & 1,19 \\
AM\,Her           & 185.7 & 0.78   & 0.15   & -     & sp     & 20\\
DW\,UMa           & 196.7 & 0.87   & 0.19   & $>$0.16 & e    & 21\\
IP\,Peg           & 227.8 & 1.16   & 0.02   & 0.55  & e      & 22\\
GY\,Cnc           & 252.6 & 0.99   & 0.12   & 0.38  & e      & 23\\
U\,Gem            & 254.7 & 1.20   & 0.09   & 0.42  & d,g,sp & 24,25,26,27,28,29\\
SDSS$1006+2337$   & 267.7 & 0.78   & 0.12   & 0.466 & e      & 30\\
DQ-Her            & 278.8 & 0.60   & 0.07   & 0.40  & d      & 31\\
V347\,Pup         & 334.0 & 0.63   & 0.04   & 0.52  & d      & 32\\
EM\,Cyg           & 418.9 & 1.00   & 0.06   & 0.77  & d      & 33\\
AC\,Cnc           & 432.7 & 0.76   & 0.03   & 0.77  & d      & 34\\
V363\,Aur         & 462.6 & 0.90   & 0.06   & 1.06  & d      & 34\\
AE\,Aqr           & 592.8 & 0.63   & 0.05   & 0.37  & d      & 35\\
\end{longtable}
\textit{References}
(1)~\citet{savouryetal11-1},
(2)~\citet{littlefairetal07-1},
(3)~\citet{pattersonetal08-1},
(4)~\citet{littlefairetal08-1},
(5)~\citet{tullochetal09-1},
(6)~\citet{steeghsetal07-1},
(7)~\citet{longetal04-1},
(8)~\citet{littlefairetal06-2},
(9)~\citet{felineetal04-2},
(10)~\citet{woodetal89-1},
(11)~\citet{felineetal04-1},
(12)~\citet{felineetal04-3},
(13)~\citet{horneetal91-1},
(14)~\citet{steeghsetal03-1},
(15)~\citet{smithetal06-1},
(16)~\citet{sionetal97-1},
(17)~\citet{wade+horne88-1},
(18)~\citet{woodetal86-1},
(19)~\citet{littlefairetal06-1},
(20)~\citet{gaensickeetal06-2},
(21)~\citet{araujo-betancoretal03-1},
(22)~\citet{copperwheatetal10-1},
(23)~\citet{thorstensen00-1},
(24)~\citet{echevarriaetal07-1},
(25)~\citet{zhangetal87-1},
(26)~\citet{sionetal98-1},
(27)~\citet{long+gilliland99-1},
(28)~\citet{nayloretal05-1},
(29)~\citet{longetal06-1},
(30)~\citet{southworthetal09-1},
(31)~\citet{horneetal93-1},
(32)~\citet{thoroughgoodetal05-1},
(33)~\citet{welshetal07-1},
(34)~\citet{thoroughgoodetal04-1},
(35)~\citet{echevarriaetal08-1}
}

\longtab{2}{
\begin{longtable}{lccccll}
\caption{\label{tab:prop} Observed properties of the PCEBs in our sample. Note that the WD
mass may change slightly (within the error) in order to be able to reconstruct the evolutionary
history of the system with $\alpha=0.25$.}\\
\hline\hline
Object    & \Mwd & $M_\mathrm{2}$ & $P_{\mathrm{obs}}$ & \Teff & Type & Ref. \\  
& [\Msun] & [\Msun] & [d] & [K] & & \\
\hline
\endfirsthead
\caption{continued.}\\
\hline\hline
Object    & \Mwd & $M_\mathrm{2}$ & $P_{\mathrm{obs}}$ & \Teff & Type & Ref. \\  
& [\Msun] & [\Msun] & [d] & [K] & & \\
\hline
\endhead
\hline
\endfoot
WD$0137-3457$   & $0.39 \pm 0.04$ & $0.05 \pm 0.01$ & $0.080$ & $16500$ & He	& 1 \\
GD\,$448$       & $0.41 \pm 0.01$ & $0.10 \pm 0.04$ & $0.103$ & $19000$ & He	& 2, 3 \\
NN\,Ser         & $0.54 \pm 0.01$ & $0.11 \pm 0.00$ & $0.130$ & $57000$ & C/O	& 4 \\
LTT\,$560$      & $0.52 \pm 0.12$ & $0.19 \pm 0.05$ & $0.148$ & $7500$  & C/O	& 5 \\
MS\,Peg         & $0.48 \pm 0.02$ & $0.22 \pm 0.02$ & $0.174$ & $22170$ & ?	& 6 \\
LM\,Com         & $0.45 \pm 0.05$ & $0.28 \pm 0.05$ & $0.259$ & $29300$ & He	& 7 \\
CC\,Cet         & $0.39 \pm 0.10$ & $0.18 \pm 0.05$ & $0.284$ & $26200$ & He	& 8 \\
CSS\,$080502$   & $0.35 \pm 0.04$ & $0.32 \pm 0.00$ & $0.149$ & $17505$ & He	& 9, 1 \\
RR\,Cae         & $0.44 \pm 0.02$ & $0.18 \pm 0.01$ & $0.303$ & $7540$  & He	& 11 \\
BPM\,$6502$     & $0.50 \pm 0.05$ & $0.17 \pm 0.01$ & $0.337$ & $21000$ & ?	& 12, 13, 14 \\
GK\,Vir         & $0.51 \pm 0.04$ & $0.10 \pm 0.00$ & $0.344$ & $48800$ & ?	& 15, 16 \\
EC\,$12477-1738$& $0.61 \pm 0.08$ & $0.38 \pm 0.07$ & $0.362$ & $17718$ & C/O	& 17 \\
EC\,$14329-1625$& $0.62 \pm 0.11$ & $0.38 \pm 0.07$ & $0.350$ & $14575$ & C/O	& 17 \\
EC\,$13349-3237$& $0.46 \pm 0.11$ & $0.50 \pm 0.05$ & $0.470$ & $35010$ & He	& 17 \\
DE\,CVn         & $0.53 \pm 0.04$ & $0.41 \pm 0.06$ & $0.364$ & $8000$  & C/O	& 18 \\
RXJ$2130.6+4710$& $0.55 \pm 0.02$ & $0.56 \pm 0.02$ & $0.521$ & $18000$ & C/O	& 19 \\
HZ\,9           & $0.51 \pm 0.10$ & $0.28 \pm 0.04$ & $0.564$ & $17400$ & ?	& 20, 21, 22, 23 \\
UX\,CVn         & $0.39 \pm 0.05$ & $0.42 \pm 0.00$ & $0.570$ & $28000$ & He	& 24 \\
UZ\,Sex         & $0.65 \pm 0.23$ & $0.22 \pm 0.05$ & $0.597$ & $19900$ & C/O	& 8, 25 \\
EG\,UMa         & $0.64 \pm 0.03$ & $0.42 \pm 0.04$ & $0.668$ & $13100$ & C/O	& 26 \\
RE\,J$2013+4002$& $0.56 \pm 0.03$ & $0.18 \pm 0.04$ & $0.706$ & $49000$ & C/O	& 27, 28 \\
RE\,J$1016-0520$& $0.60 \pm 0.02$ & $0.15 \pm 0.02$ & $0.789$ & $55000$ & C/O	& 27, 28 \\
IN\,CMa         & $0.57 \pm 0.03$ & $0.43 \pm 0.03$ & $1.260$ & $52400$ & C/O	& 27, 29 \\
Feige\,24       & $0.57 \pm 0.03$ & $0.39 \pm 0.02$ & $4.232$ & $57000$ & C/O	& 30 \\
IK\,Peg         & $1.19 \pm 0.05$ & $1.70 \pm 0.10$ & $21.722$& $35500$ & C/O	& 29, 31 \\
SDSS$0052-0053$ & $1.22 \pm 0.17$ & $0.32 \pm 0.06$ & $0.114$ & $16108$ & C/O	& 32 \\
SDSS$0110+1326$ & $0.47 \pm 0.02$ & $0.32 \pm 0.05$ & $0.333$ & $25891$ & He	& 10 \\
SDSS$0238-0005$ & $0.48 \pm 0.15$ & $0.38 \pm 0.01$ & $0.212$ & $20566$ & ?	& 33 \\
SDSS$0246+0041$ & $0.80 \pm 0.07$ & $0.38 \pm 0.01$ & $0.728$ & $17452$ & C/O	& 32 \\
SDSS$0303+0054$ & $0.92 \pm 0.04$ & $0.25 \pm 0.03$ & $0.134$ & $\sim 8000$ & C/O & 10 \\
SDSS$0833+0702$ & $0.54 \pm 0.07$ & $0.32 \pm 0.06$ & $0.304$ & $15246$ & C/O	& 33 \\
SDSS$0924+0024$ & $0.52 \pm 0.03$ & $0.32 \pm 0.06$ & $2.404$ & $19193$ & C/O	& 33 \\
SDSS$0949+0322$ & $0.51 \pm 0.08$ & $0.32 \pm 0.06$ & $0.396$ & $18542$ & ?	& 33 \\
SDSS$1047+0523$ & $0.38 \pm 0.17$ & $0.26 \pm 0.04$ & $0.382$ & $12392$ & He	& 34 \\
SDSS$1143+0009$ & $0.60 \pm 0.04$ & $0.32 \pm 0.06$ & $0.386$ & $16910$ & C/O	& 33 \\
SDSS$1212-0123$ & $0.47 \pm 0.01$ & $0.28 \pm 0.02$ & $0.336$ & $17700$ & He	& 35 \\
SDSS$1348+1834$ & $0.59 \pm 0.02$ & $0.32 \pm 0.06$ & $0.248$ & $15071$ & C/O	& 33 \\
SDSS$1411+1028$ & $0.54 \pm 0.08$ & $0.38 \pm 0.01$ & $0.167$ & $30419$ & C/O	& 33 \\
SDSS$1414-0132$ & $0.67 \pm 0.15$ & $0.26 \pm 0.04$ & $0.728$ & $13588$ & C/O	& 34 \\
SDSS$1429+5759$ & $1.07 \pm 0.13$ & $0.38 \pm 0.01$ & $0.545$ & $16149$ & C/O	& 33 \\
SDSS$1434+5335$ & $0.49 \pm 0.02$ & $0.32 \pm 0.06$ & $4.357$ & $21785$ & ?	& 33 \\
SDSS$1435+3733$ & $0.42 \pm 0.05$ & $0.26 \pm 0.04$ & $0.126$ & $12392$ & He	& 10 \\
SDSS$1506-0120$ & $0.45 \pm 0.09$ & $0.32 \pm 0.06$ & $1.051$ & $15601$ & He	& 33 \\
SDSS$1519+3536$ & $0.57 \pm 0.03$ & $0.20 \pm 0.04$ & $1.367$ & $19416$ & C/O	& 33 \\
SDSS$1524+5040$ & $0.73 \pm 0.03$ & $0.32 \pm 0.06$ & $0.590$ & $20098$ & C/O	& 33 \\
SDSS$1529+0020$ & $0.39 \pm 0.02$ & $0.26 \pm 0.04$ & $0.165$ & $13986$ & He	& 32 \\
SDSS$1548+4057$ & $0.62 \pm 0.13$ & $0.20 \pm 0.04$ & $0.185$ & $11699$ & C/O	& 10 \\
SDSS$1558+2642$ & $1.06 \pm 0.31$ & $0.32 \pm 0.06$ & $0.661$ & $14560$ & C/O	& 33 \\
SDSS$1646+4223$ & $0.53 \pm 0.06$ & $0.26 \pm 0.04$ & $1.595$ & $17707$ & C/O	& 33 \\
SDSS$1705+2109$ & $0.52 \pm 0.04$ & $0.26 \pm 0.04$ & $0.815$ & $23886$ & C/O	& 33 \\
SDSS$1718+6101$ & $0.53 \pm 0.06$ & $0.32 \pm 0.06$ & $0.673$ & $18120$ & C/O	& 33 \\
SDSS$1724+5620$ & $0.42 \pm 0.01$ & $0.36 \pm 0.07$ & $0.333$ & $35746$ & He	& 32 \\
SDSS$1731+6233$ & $0.34 \pm 0.04$ & $0.32 \pm 0.06$ & $0.268$ & $15601$ & He	& 33 \\
SDSS$2112+1014$ & $1.06 \pm 0.05$ & $0.20 \pm 0.04$ & $0.092$ & $19868$ & C/O	& 33 \\
SDSS$2114-0103$ & $0.70 \pm 0.07$ & $0.38 \pm 0.01$ & $0.411$ & $28064$ & C/O	& 33 \\
SDSS$2120-0058$ & $0.64 \pm 0.04$ & $0.32 \pm 0.06$ & $0.449$ & $16336$ & C/O	& 33 \\
SDSS$2123+0024$ & $0.31 \pm 0.07$ & $0.20 \pm 0.04$ & $0.149$ & $13432$ & He	& 33 \\
SDSS$2132+0031$ & $0.39 \pm 0.03$ & $0.32 \pm 0.06$ & $0.222$ & $16336$ & He	& 33 \\
SDSS$2216+0102$ & $0.41 \pm 0.14$ & $0.26 \pm 0.04$ & $0.210$ & $12536$ & He	& 33 \\
SDSS$2240-0935$ & $0.41 \pm 0.07$ & $0.26 \pm 0.04$ & $0.261$ & $12536$ & He	& 33 \\
SDSS$2318-0935$ & $0.50 \pm 0.05$ & $0.38 \pm 0.01$ & $2.534$ & $22550$ & ?	& 33 \\
SDSS$2339-0020$ & $0.93 \pm 0.18$ & $0.32 \pm 0.06$ & $0.655$ & $14560$ & C/O	& 32 \\
\end{longtable}
\textit{References}.
(1) \citet{burleighetal06-1},
(2) \citet{marsh+duck96-1},
(3) \citet{maxtedetal98-1},
(4) \citet{parsonsetal10-1},
(5) \citet{tappertetal07-1},
(6) \citet{schmidtetal95-3},
(7) \citet{oroszetal99-1},
(8) \citet{safferetal93-1},
(9) \citet{drakeetal09-1},
(10) \citet{pyrzasetal09-1},
(11) \citet{maxtedetal07-2},
(12) \citet{kawkaetal00-1},
(13) \citet{bragagliaetal95-1},
(14) \citet{koesteretal79-1},
(15) \citet{fulbrightetal93-1},
(16) \citet{greenetal78-1},
(17) \citet{tappertetal09-1},
(18) \citet{vandenbesselaaretal07-1},
(19) \citet{maxtedetal04-1},
(20) \citet{stauffer87-1},
(21) \citet{lanning+pesch81-1},
(22) \citet{guinan+sion84-1},
(23) \citet{schreiber+gaensicke03-1},
(24) \citet{Hillwigetal00-1},
(25) \citet{kepler+nelan93-1}
(26) \citet{bleachetal00-1},
(27) \citet{vennesetal99-2},
(28) \citet{bergeronetal94-1},
(29) \citet{davisetal10-1},
(30)\citet{kawkaetal08-1},
(31) \citet{landsmanetal93-1},
(32) \citet{rebassa-mansergasetal08-1}, 
(33) Nebot G\'omez-Mor\'an et al. (2011, submitted),
(34) \citet{schreiberetal08-1},
(35) \citet{nebot-gomez-moranetal09-1}.
}

\longtab{3}{
\begin{landscape}
\begin{longtable}{lcccccccccccl}
\caption{\label{tab:res} Derived properties of the PCEBs in our sample.}\\
\hline\hline
Object    & \Mwd & q & $P_{\mathrm{CE}}$ & $M_{\mathrm{1,o}}$ & $M_{\mathrm{1,CE}}$ & $a_{\mathrm{i}}$ &  $P_{\mathrm{sd}}$ & $t_{\mathrm{evolv}}$ & $t_{\mathrm{cool}}$ & $t_{\mathrm{sd}}$ & $t_{\mathrm{tot}}$ & Notes\\
& [\Msun] &  & [d] & [\Msun] & [\Msun] & [\Rsun] &  [h] & [yr] & [yr] & [yr] & [yr] &\\
\hline
\endfirsthead
\caption{continued.}\\
\hline\hline
Object    & \Mwd & q & $P_{\mathrm{CE}}$ & $M_{\mathrm{1,o}}$ & $M_{\mathrm{1,CE}}$ & $a_{\mathrm{i}}$ &  $P_{\mathrm{sd}}$ & $t_{\mathrm{evolv}}$ & $t_{\mathrm{cool}}$ & $t_{\mathrm{sd}}$ & $t_{\mathrm{tot}}$ & Notes\\
& [\Msun] &  & [d] & [\Msun] & [\Msun] & [\Rsun] &  [h] & [yr] & [yr] & [yr] & [yr] &\\
\hline
\endhead
\hline
\endfoot
WD$0137-3457$   &  $0.39$ &  $0.14$ &  $0.083$ &  $0.980$ &  $0.900$ &  $124.06$ &  $0.921$ &  $1.33E+10$ &  $2.02E+08$ &  $1.65E+09$ &  $1.51E+10$ &  c \\
\textbf{GD\,448}  &  \textbf{0.41} &  \textbf{0.23} &  \textbf{0.105} &  \textbf{1.190} &  \textbf{1.120} &  \textbf{152.23} &  \textbf{1.282} &  \textbf{6.57E+09} &  \textbf{1.18E+08} &  \textbf{1.84E+09} &  \textbf{8.53E+09} &  \textbf{a} \\
\textbf{NN\,Ser } &  \textbf{0.54} &  \textbf{0.21} &  \textbf{0.130} &  \textbf{2.010} &  \textbf{1.964} &  \textbf{287.58} &  \textbf{1.825} &  \textbf{1.47E+09} &  \textbf{1.18E+06} &  \textbf{2.26E+09} &  \textbf{3.73E+09} &  \textbf{a} \\
\textbf{LTT\,560} &  \textbf{0.52} &  \textbf{0.36} &  \textbf{0.168} &  \textbf{1.690} &  \textbf{1.615} &  \textbf{251.56} &  \textbf{2.204} &  \textbf{2.15E+09} &  \textbf{1.04E+09} &  \textbf{1.88E+09} &  \textbf{5.07E+09} &  \textbf{a} \\
\textbf{MS\,Peg}  &  \textbf{0.46} &  \textbf{0.48} &  \textbf{0.174} &  \textbf{1.750} &  \textbf{1.725} &  \textbf{225.23} &  \textbf{2.651} &  \textbf{1.80E+09} &  \textbf{7.82E+07} &  \textbf{2.67E+09} &  \textbf{4.55E+09} &  \textbf{a} \\
\textbf{LM\,Com}  &  \textbf{0.45} &  \textbf{0.62} &  \textbf{0.259} &  \textbf{1.670} &  \textbf{1.641} &  \textbf{224.19} &  \textbf{2.660} &  \textbf{2.09E+09} &  \textbf{3.24E+07} &  \textbf{8.23E+09} &  \textbf{1.04E+10} &  \textbf{a} \\
CC\,Cet         &  $0.39$ &  $0.46$ &  $0.284$ &  $1.020$ &  $0.945$ &  $144.91$ &  $2.016$ &  $1.15E+10$ &  $4.28E+07$ &  $1.78E+10$ &  $2.93E+10$ &  c \\
CSS\,080502     &  $0.35$ &  $0.91$ &  $0.288$ &  $0.990$ &  $0.945$ &  $103.79$ &  $2.873$ &  $1.28E+10$ &  $1.45E+08$ &  $9.13E+06$ &  $1.29E+10$ &  b \\
RR\,Cae         &  $0.44$ &  $0.41$ &  $0.314$ &  $1.310$ &  $1.240$ &  $213.61$ &  $1.715$ &  $4.68E+09$ &  $2.04E+09$ &  $2.04E+10$ &  $2.71E+10$ &  c \\
BPM\,6502       &  $0.52$ &  $0.33$ &  $0.337$ &  $1.410$ &  $1.276$ &  $287.03$ &  $2.001$ &  $3.78E+09$ &  $3.79E+07$ &  $2.56E+10$ &  $2.94E+10$ &  c \\
GK\,Vir         &  $0.52$ &  $0.19$ &  $0.344$ &  $1.510$ &  $1.387$ &  $345.24$ &  $1.652$ &  $3.03E+09$ &  $2.02E+06$ &  $4.48E+10$ &  $4.78E+10$ &  c \\
\textbf{EC\,12477-1738}&  \textbf{0.61} &  \textbf{0.62} &  \textbf{0.386} &  \textbf{2.590} &  \textbf{2.525} &  \textbf{372.82} &  \textbf{3.467} &  \textbf{7.24E+08} &  \textbf{1.13E+08} &  \textbf{4.42E+08} &  \textbf{1.28E+09} &  \textbf{a} \\
\textbf{EC\,14329-1625}&  \textbf{0.62} &  \textbf{0.61} &  \textbf{0.397} &  \textbf{2.650} &  \textbf{2.583} &  \textbf{383.44} &  \textbf{3.471} &  \textbf{6.78E+08} &  \textbf{2.20E+08} &  \textbf{3.98E+08} &  \textbf{1.34E+09} &  \textbf{a} \\
EC\,$13349-3237$&  $0.46$ &  $1.09$ &  $0.470$ &  $1.750$ &  $1.725$ &  $257.59$ &  $3.938$ &  $1.80E+09$ &  $<5.0E07 $ &  $1.03E+09$ &  $2.83E+09$ &  b \\
DE\,CVn         &  $0.53$ &  $0.77$ &  $0.477$ &  $1.860$ &  $1.810$ &  $297.61$ &  $3.028$ &  $1.65E+09$ &  $8.95E+08$ &  $5.97E+08$ &  $3.14E+09$ &  b \\
RXJ$2130.6+4710$&  $0.55$ &  $1.01$ &  $0.529$ &  $2.140$ &  $2.099$ &  $312.03$ &  $4.417$ &  $1.24E+09$ &  $8.84E+07$ &  $1.73E+09$ &  $3.06E+09$ &  b \\
HZ\,9           &  $0.52$ &  $0.54$ &  $0.564$ &  $1.440$ &  $1.314$ &  $309.08$ &  $2.679$ &  $3.53E+09$ &  $8.71E+07$ &  $6.52E+10$ &  $6.88E+10$ &  c \\
UX\,CVn         &  $0.39$ &  $1.08$ &  $0.570$ &  $1.110$ &  $1.043$ &  $162.31$ &  $3.237$ &  $8.44E+09$ &  $<5.0E07 $ &  $1.86E+09$ &  $1.03E+10$ &  b \\
UZ\,Sex         &  $0.65$ &  $0.34$ &  $0.597$ &  $2.710$ &  $2.586$ &  $451.50$ &  $2.385$ &  $6.37E+08$ &  $8.37E+07$ &  $7.88E+10$ &  $7.96E+10$ &  c \\
\textbf{EG\,UMa}  &  \textbf{0.64} &  \textbf{0.66} &  \textbf{0.683} &  \textbf{2.730} &  \textbf{2.641} &  \textbf{439.15} &  \textbf{3.659} &  \textbf{6.23E+08} &  \textbf{3.13E+08} &  \textbf{3.94E+09} &  \textbf{4.88E+09} &  \textbf{a} \\
RE\,J$2013+4002$&  $0.56$ &  $0.32$ &  $0.706$ &  $2.170$ &  $2.102$ &  $391.24$ &  $2.194$ &  $1.19E+09$ &  $2.10E+06$ &  $1.66E+11$ &  $1.68E+11$ &  c \\
RE\,J$1016-0520$&  $0.60$ &  $0.25$ &  $0.789$ &  $2.360$ &  $2.243$ &  $423.38$ &  $1.998$ &  $9.41E+08$ &  $1.60E+06$ &  $2.52E+11$ &  $2.53E+11$ &  c \\
IN\,CMa         &  $0.57$ &  $0.75$ &  $1.260$ &  $2.270$ &  $2.202$ &  $422.04$ &  $3.023$ &  $1.05E+09$ &  $1.73E+06$ &  $4.51E+10$ &  $4.61E+10$ &  b,c\\
Feige\,24       &  $0.57$ &  $0.68$ &  $4.232$ &  $2.190$ &  $2.100$ &  $474.80$ &  $3.195$ &  $1.16E+09$ &  $1.33E+06$ &  $1.94E+12$ &  $1.94E+12$ &  b,c\\
IK\,Peg         &  $1.19$ &  $1.43$ &  $21.722$&  $6.500$ &  $6.124$ &  $1300.58$& $521.32$ &  $6.55E+07$ &  $2.70E+07$ &  $1.88E+09$ &  $1.97E+09$ &  * \\
SDSS$0052-0053$ &  $1.22$ &  $0.26$ &  $0.297$ &  $6.740$ &  $6.464$ &  $873.70$ &  $2.960$ &  $6.03E+07$ &  $4.22E+08$ &  $-5.32E+06$&  $4.77E+08$ &  d \\
SDSS$0110+1326$ &  $0.46$ &  $0.69$ &  $0.347$ &  $1.720$ &  $1.690$ &  $254.65$ &  $2.865$ &  $1.91E+09$ &  $4.61E+07$ &  $3.01E+08$ &  $2.25E+09$ &  b \\
SDSS$0238-0005$ &  $0.47$ &  $0.79$ &  $0.243$ &  $1.730$ &  $1.702$ &  $258.63$ &  $3.412$ &  $1.87E+09$ &  $3.70E+07$ &  $4.69E+07$ &  $1.95E+09$ &  b \\
\textbf{SDSS0246+0041} &  \textbf{0.80} &  \textbf{0.47} &  \textbf{0.735} &  \textbf{3.590} &  \textbf{3.519} &  \textbf{573.02} &  \textbf{3.514} &  \textbf{2.88E+08} &  \textbf{2.03E+08} &  \textbf{5.85E+09} &  \textbf{6.34E+09} &  \textbf{a} \\
\textbf{SDSS0303+0054} &  \textbf{0.92} &  \textbf{0.27} &  \textbf{0.207} &  \textbf{4.600} &  \textbf{4.505} &  \textbf{614.05} &  \textbf{2.840} &  \textbf{1.49E+08} &  \textbf{2.25E+09} &  \textbf{2.88E+08} &  \textbf{2.68E+09} &  \textbf{a} \\
\textbf{SDSS0833+0702} &  \textbf{0.54} &  \textbf{0.59} &  \textbf{0.351} &  \textbf{2.010} &  \textbf{1.968} &  \textbf{290.08} &  \textbf{2.889} &  \textbf{1.47E+09} &  \textbf{1.57E+08} &  \textbf{2.45E+08} &  \textbf{1.88E+09} &  \textbf{a} \\
SDSS$0924+0024$ &  $0.52$ &  $0.62$ &  $2.404$ &  $1.470$ &  $1.324$ &  $425.88$ &  $2.881$ &  $3.30E+09$ &  $5.74E+07$ &  $2.43E+11$ &  $2.47E+11$ &  c \\
\textbf{SDSS0949+0322} &  \textbf{0.52} &  \textbf{0.62} &  \textbf{0.408} &  \textbf{1.600} &  \textbf{1.509} &  \textbf{291.89} &  \textbf{2.880} &  \textbf{2.53E+09} &  \textbf{6.65E+07} &  \textbf{5.83E+08} &  \textbf{3.18E+09} &  \textbf{a} \\
SDSS$1047+0523$ &  $0.38$ &  $0.68$ &  $0.384$ &  $1.000$ &  $0.932$ &  $140.19$ &  $2.253$ &  $1.23E+10$ &  $4.58E+08$ &  $2.90E+10$ &  $4.18E+10$ &  b,c\\
\textbf{SDSS1143+0009} &  \textbf{0.60} &  \textbf{0.53} &  \textbf{0.409} &  \textbf{2.470} &  \textbf{2.393} &  \textbf{379.70} &  \textbf{2.905} &  \textbf{8.27E+08} &  \textbf{1.29E+08} &  \textbf{6.07E+08} &  \textbf{1.56E+09} &  \textbf{a} \\
SDSS$1212-0123$ &  $0.47$ &  $0.59$ &  $0.337$ &  $1.690$ &  $1.657$ &  $261.49$ &  $3.118$ &  $2.01E+09$ &  $1.76E+08$ &  $1.66E+10$ &  $1.88E+10$ &  c \\
\textbf{SDSS1348+1834} &  \textbf{0.59} &  \textbf{0.54} &  \textbf{0.319} &  \textbf{2.460} &  \textbf{2.400} &  \textbf{352.99} &  \textbf{2.902} &  \textbf{8.37E+08} &  \textbf{1.84E+08} &  \textbf{1.27E+08} &  \textbf{1.15E+09} &  \textbf{a} \\
SDSS$1411+1028$ &  $0.54$ &  $0.70$ &  $0.180$ &  $2.010$ &  $1.975$ &  $202.63$ &  $3.441$ &  $1.47E+09$ &  $8.87E+06$ &  $1.29E+07$ &  $1.49E+09$ &  b \\
SDSS$1414-0132$ &  $0.67$ &  $0.39$ &  $0.729$ &  $2.850$ &  $2.728$ &  $480.64$ &  $2.301$ &  $5.52E+08$ &  $3.05E+08$ &  $1.13E+11$ &  $1.14E+11$ &  c \\
\textbf{SDSS1429+5759} &  \textbf{1.07} &  \textbf{0.36} &  \textbf{0.568} &  \textbf{5.700} &  \textbf{5.517} &  \textbf{810.37} &  \textbf{3.547} &  \textbf{8.83E+07} &  \textbf{4.18E+08} &  \textbf{2.76E+09} &  \textbf{3.27E+09} &  \textbf{a} \\
SDSS$1434+5335$ &  $0.52$ &  $0.62$ &  $4.357$ &  $1.320$ &  $1.137$ &  $439.49$ &  $2.880$ &  $4.70E+09$ &  $3.21E+07$ &  $1.76E+12$ &  $1.76E+12$ &  c \\
\textbf{SDSS1435+3733} &  \textbf{0.42} &  \textbf{0.62} &  \textbf{0.140} &  \textbf{1.580} &  \textbf{1.554} &  \textbf{169.85} &  \textbf{2.256} &  \textbf{2.50E+09} &  \textbf{4.73E+08} &  \textbf{8.19E+08} &  \textbf{3.79E+09} &  \textbf{a} \\
SDSS$1506-0120$ &  $0.45$ &  $0.71$ &  $1.057$ &  $1.330$ &  $1.255$ &  $257.22$ &  $2.856$ &  $4.44E+09$ &  $2.56E+08$ &  $1.38E+10$ &  $1.85E+10$ &  b,c\\
SDSS$1519+3536$ &  $0.57$ &  $0.35$ &  $1.367$ &  $2.200$ &  $2.113$ &  $422.68$ &  $1.733$ &  $1.15E+09$ &  $6.44E+07$ &  $1.26E+12$ &  $1.26E+12$ &  c \\
\textbf{SDSS1524+5040} &  \textbf{0.73} &  \textbf{0.44} &  \textbf{0.596} &  \textbf{3.240} &  \textbf{3.142} &  \textbf{531.79} &  \textbf{2.928} &  \textbf{3.83E+08} &  \textbf{1.07E+08} &  \textbf{2.94E+09} &  \textbf{3.43E+09} &  \textbf{a} \\
SDSS$1529+0020$ &  $0.39$ &  $0.67$ &  $0.170$ &  $1.270$ &  $1.228$ &  $137.92$ &  $2.245$ &  $5.22E+09$ &  $2.98E+08$ &  $2.59E+09$ &  $8.10E+09$ &  b \\
\textbf{SDSS1548+4057} &  \textbf{0.62} &  \textbf{0.32} &  \textbf{0.192} &  \textbf{2.630} &  \textbf{2.556} &  \textbf{359.75} &  \textbf{1.736} &  \textbf{6.93E+08} &  \textbf{3.98E+08} &  \textbf{3.64E+09} &  \textbf{4.74E+09} &  \textbf{a} \\
\textbf{SDSS1558+2642} &  \textbf{1.06} &  \textbf{0.30} &  \textbf{0.681} &  \textbf{5.620} &  \textbf{5.418} &  \textbf{824.56} &  \textbf{2.956} &  \textbf{9.13E+07} &  \textbf{6.09E+08} &  \textbf{5.75E+09} &  \textbf{6.45E+09} &  \textbf{a} \\
SDSS$1646+4223$ &  $0.53$ &  $0.49$ &  $1.595$ &  $1.570$ &  $1.419$ &  $422.70$ &  $2.282$ &  $2.69E+09$ &  $8.68E+07$ &  $1.10E+12$ &  $1.10E+12$ &  c \\
SDSS$1705+2109$ &  $0.52$ &  $0.50$ &  $0.815$ &  $1.630$ &  $1.532$ &  $377.68$ &  $2.281$ &  $2.40E+09$ &  $2.20E+07$ &  $1.86E+11$ &  $1.88E+11$ &  c \\
\textbf{SDSS1718+6101} &  \textbf{0.53} &  \textbf{0.60} &  \textbf{0.678} &  \textbf{1.860} &  \textbf{1.802} &  \textbf{359.97} &  \textbf{2.886} &  \textbf{1.65E+09} &  \textbf{7.86E+07} &  \textbf{3.56E+09} &  \textbf{5.29E+09} &  \textbf{a} \\
SDSS$1724+5620$ &  $0.42$ &  $0.86$ &  $0.333$ &  $1.410$ &  $1.368$ &  $190.02$ &  $3.483$ &  $3.64E+09$ &  $<5.0E07 $ &  $2.10E+08$ &  $3.85E+09$ &  b \\
SDSS$1731+6233$ &  $0.34$ &  $0.94$ &  $0.268$ &  $1.000$ &  $0.962$ &  $90.95$ &  $2.794$ &  $1.23E+10$ &  $<5.0E07 $ &  $1.09E+08$ &  $1.24E+09$ &  b \\
\textbf{SDSS2112+1014} &  \textbf{1.06} &  \textbf{0.19} &  \textbf{0.109} &  \textbf{5.650} &  \textbf{5.516} &  \textbf{682.31} &  \textbf{1.742} &  \textbf{9.00E+07} &  \textbf{2.29E+08} &  \textbf{1.98E+08} &  \textbf{5.17E+08} &  \textbf{a} \\
\textbf{SDSS2114-0103} &  \textbf{0.70} &  \textbf{0.54} &  \textbf{0.413} &  \textbf{3.130} &  \textbf{3.063} &  \textbf{463.72} &  \textbf{3.493} &  \textbf{4.23E+08} &  \textbf{1.66E+07} &  \textbf{7.62E+08} &  \textbf{1.20E+09} &  \textbf{a} \\
\textbf{SDSS2120-0058} &  \textbf{0.64} &  \textbf{0.50} &  \textbf{0.469} &  \textbf{2.710} &  \textbf{2.615} &  \textbf{427.85} &  \textbf{2.913} &  \textbf{6.37E+08} &  \textbf{1.65E+08} &  \textbf{1.06E+09} &  \textbf{1.86E+09} &  \textbf{a} \\
SDSS$2123+0024$ &  $0.31$ &  $0.65$ &  $0.149$ &  $0.980$ &  $0.946$ &  $75.72$ &  $1.690$ &  $1.33E+10$ &  $<5.0E07 $ &  $3.30E+09$ &  $1.66E+10$ &  c \\
SDSS$2132+0031$ &  $0.39$ &  $0.82$ &  $0.328$ &  $1.160$ &  $1.101$ &  $149.49$ &  $2.826$ &  $7.20E+09$ &  $1.77E+08$ &  $6.06E+07$ &  $7.44E+09$ &  b \\
SDSS$2216+0102$ &  $0.41$ &  $0.63$ &  $0.216$ &  $1.150$ &  $1.071$ &  $180.10$ &  $1.484$ &  $7.43E+09$ &  $4.43E+08$ &  $5.83E+09$ &  $1.37E+10$ &  c \\
SDSS$2240-0935$ &  $0.41$ &  $0.63$ &  $0.265$ &  $1.280$ &  $1.227$ &  $169.38$ &  $2.254$ &  $5.08E+09$ &  $4.47E+08$ &  $1.00E+10$ &  $1.55E+10$ &  c \\
SDSS$2318-0935$ &  $0.52$ &  $0.74$ &  $2.534$ &  $1.450$ &  $1.311$ &  $430.92$ &  $3.428$ &  $3.45E+09$ &  $2.74E+07$ &  $2.64E+11$ &  $2.68E+11$ &  b,c\\
\textbf{SDSS2339-0020} &  \textbf{0.93} &  \textbf{0.34} &  \textbf{0.674} &  \textbf{4.660} &  \textbf{4.526} &  \textbf{706.93} &  \textbf{2.948} &  \textbf{1.44E+08} &  \textbf{5.08E+08} &  \textbf{5.00E+09} &  \textbf{5.66E+09} &  \textbf{a} \\
\end{longtable}
*For IK\,Peg the values of $P_{\mathrm{sd}}$ and $t_{\mathrm{sd}}$ 
were computed based on the nuclear evolution timescale of the secondary as it is much shorter\\ 
than the time required to bring the systems into contact by angular momentum loss.\\ 
Notes:\\
a) The system is representative for the progenitors of the current CV population.\\
b) The mass ratio exceeds the critical value for stable mass transfer.\\
c) The system has a CV formation times longer than the age of the galaxy.\\
d) This system is probably not a PCEB but a detached CV in the period gap. \\
\end{landscape}
}

\end{document}